\documentclass[reprint]{revtex4-1}
\usepackage{amsmath,amssymb,amsfonts} 
\usepackage[colorlinks,citecolor=blue,urlcolor=blue]{hyperref}
\usepackage[pdftex]{graphicx}
\usepackage{url}
\usepackage[tight,footnotesize]{subfigure}
\DeclareMathOperator*{\argmin}{arg\,min}

\usepackage{algcompatible}
\usepackage{newfloat}

\DeclareFloatingEnvironment[
    fileext=loa,
    listname=List of Algorithms,
    name=ALGORITHM,
    placement=tbhp,
]{algorithm}

\begin{document}
\title{Structural and Functional Discovery in Dynamic Networks\\with Non-negative Matrix Factorization}
\author{Shawn Mankad}
\author{George Michailidis}
\affiliation{Statistics Department, University of Michigan}
\begin{abstract}
Time series of graphs are increasingly prevalent in modern data and pose unique challenges to visual exploration and pattern extraction. This paper describes the development and application of matrix factorizations for exploration and time-varying community detection in time-evolving graph sequences. The matrix factorization model allows the user to home in on and display interesting, underlying structure and its evolution over time. The methods are scalable to weighted networks with a large number of time points or nodes, and can accommodate sudden changes to graph topology. Our techniques are demonstrated with several dynamic graph series from both synthetic and real world data, including citation and trade networks. These examples illustrate how users can steer the techniques and combine them with existing methods to discover and display meaningful patterns in sizable graphs over many time points.
\end{abstract}
\maketitle

\section{Introduction}
Due to advances in data collection
technologies, it is becoming increasingly common to study
time series of networks. An important research question is 
how to discover the underlying structure and 
dynamics in time-varying networked systems. In this work, we propose a 
new matrix factorization-based approach for community discovery and 
visual exploration within potentially weighted and directed network time-series. 
Next, we review and discuss this work in relation to 
popular approaches for addressing the key problems of 
community detection and visualization of time series of networks.

There have been many important 
contributions for community detection
in network time-series, extensively reviewed in 
\cite{Fienberg-networkOverview-jcgs,2009arXiv0912.5410G}, 
from the fields of physics, computer science and statistics. 
The basic goal of community detection 
is to extract groups of nodes that feature relatively dense
within group connectivity and sparser between group connections
\cite{Newman2,PhysRevE.84.036103}. 
A common strategy is to embed the graphs in low-dimensional latent spaces. 
For instance,
\cite{shedden-newman} use 
latent variables to capture groups of papers that evolve similarly in
citation network data. 
\cite{sarkar-moore} extend to the dynamic setting a popular latent space model 
for static data \cite{hoff-network-latentspace} by utilizing 
smoothness constraints to preserve the coordinates of the nodes in the latent space over time. 
This article also utilizes a similar low-dimensional embedding strategy. 
A key difference between this work and \cite{sarkar-moore} is that community 
membership itself is subject to smoothness conditions in our approach, 
hence removing the need for a two stage procedure.

This article is also in contrast to previous works that 
use temporal smoothness constraints for 
non-overlapping (hard) community detection \cite{Sun:2007:GPM:1281192.1281266}, 
estimating time-varying network structure from covariate information \cite{kolarTimeNetworks},  
predicting network (link) structure \cite{2012arXiv1205.1406R}, or anomaly detection \cite{Asur:2009:EFC:1631162.1631164,6208875}.

A sequence of non-negative factorizations 
discovers overlapping community structure, where 
node participation within each community is quantified and time-varying. 
Other works that consider 
a single network cross-section have 
shown advantages of NMF for community detection \cite{bayesNMF-networks, Ding2}.
In addition to a  
quantification of how strongly each node participates in each community, 
NMF does not suffer from the drawbacks of modularity optimization methods, 
such as the resolution limit \cite{Fortunato02012007}. 

We also use the NMF to transform 
the time series of networks to a time-series for each node, which can be 
used to create an alternative to graph drawings for visualization of 
node dynamics. Much of the visualization literature aims to enhance 
static graph drawing methods with animations that move nodes (vertices) as little
as possible between time steps to facilitate
readability \cite{onlineDrawing}.  However, the reliability of these
methods rely on the human ability to perceive and remember changes
\cite{mentalMap}. Moreover, experiments have discovered that the
effectiveness of dynamic layouts are strongly predicted by node speed
and target separation \cite{Ghani2012}. Thus, dynamic graph drawings
encounter difficulties when faced with a large number of time points,
larger graphs that feature abrupt, non-smooth changes, or if the user
is interested in detailed analysis, especially at the individual node
level (see Section 3.2 of \cite{Landesberger},
\cite{Landesberger2,TimeMatrix}). 
On the other hand, static displays facilitate detailed analysis and avoid
difficulties associated with animated layouts. This highlights a 
main advantage our NMF model, namely creating static displays of node evolutions.

The remainder of this article is organized as follows: in the next
section, we introduce a model for static network data in Section
\ref{sec:static}, followed by an extension for dynamic networks in
Section \ref{sec:overview}. We
then test the matrix factorization model on several synthetic and
real-world data sets in Section \ref{sec:experiments}. In Section
\ref{sec:discussion}, we close the article with a brief discussion.

\section{NMF for Network Cross-sections} \label{sec:static}
The most common factorization is the Singular Value Decomposition
(SVD), which has important connections to community detection, graph
drawing, and areas of statistics and signal processing
\cite{Hast:Tibs:Frie:2001}. For instance in classical spectral layout,
the coordinates of each node are given by the SVD of graph related
matrices, and can be calculated efficiently using algorithms in
\cite{GraphDrawing-Koren,Dynamic-Spectral-Drawing}.  Recently, there
has been extensive interest in spectral clustering
\cite{2012arXiv1204.2296R, 1227.62042,Chung_Laplacian}, which aims to
discover community structure in eigenvectors of the graph Laplacian
matrix. The method proposed in this paper is similar in spirit, as it
also relies on low rank approximations to adjacency matrices 
(instead of Laplacian matrices).  However, we
search for low-rank approximations that satisfy different (relaxed) constraints
than orthonormality, namely, that the approximating
decompositions are composed of non-negative entries. Such
factorizations, referred to as NMF, have been shown to be advantageous
for visualization of non-negative data
\cite{NNMF1,NNMF2,ALS,plos-nmf-biology}.  Non-negativity is typically
satisfied with networks, as edges commonly correspond to flows,
capacity, or binary relationships, and hence are non-negative.  NMF
solutions do not have simple expressions in terms of eigenvectors.
They can, however, be efficiently computed by formulating the problem
as one of penalized optimization, and using modern gradient-descent
algorithms.  Recently, theoretical connections between NMF and
important problems in data mining have been
developed \cite{Ding,NMF-PLSI}, and accordingly, NMF has been proposed
for overlapping community detection on static
\cite{bayesNMF-networks,Ding2} and dynamic \cite{facetnet} networks .

With NMF a given adjacency matrix is approximated
with an outer product that is estimated through the following minimization
\begin{equation}\label{obj:simple}
\min_{U \ge 0, V \ge 0} ||A - U V^T ||_{F}^2,
\end{equation}
where $A$ is the $n\times n$ adjacency matrix, and $U$ and $V$ are
both $n\times K$ matrices with elements in $\mathbb{R}_{+}$. The rank
or dimension of the approximation $K$ corresponds to the number of
communities, and is chosen to obtain a good fit to the data while
achieving interpretability. An interesting fact about NMF is that 
the estimates are always rescalable (scale invariant). 
For example, we can multiply $U$ by some constant $c$ and $V$ by $1/c$ 
to obtain different $U,V$ estimates without changing their product $UV^{T}$. 
Thus, as seen by the rotational indeterminancy and multiplicative nature 
of the factorization, NMF is an under-constrained model.

It is, however, straightforward to interpret the estimates due to
non-negativity. For instance, $(U)_{ik}(V)_{jk}$ can be interpreted as
the contribution of the $k$th cluster to the edge $(A)_{ij}$.
In other words, the expected interaction 
$(\hat{A})_{ij} = \sum_{k=1}^{K} (U)_{ik}(V)_{jk}$ between nodes $i$
and $j$ is the result of their mutual participation in 
the same communities \cite{bayesNMF-networks}. 
Such an edge decomposition can then be used to 
assign nodes to communities. For instance, one can proceed by 
first assigning all edges to the community with largest
relative contribution. Then, nodes are assigned to communities 
according to the proportion of its edges that belong to each
community. 
We note that with an NMF-based methodology, the adjacency matrix can be 
weighted (non-negatively), a potentially appealing feature since 
many existing analysis tools are arguably only compatible with 
networks of binary relations.

Though it is not explicitly controlled, standard NMF tends to estimate
sparse components. Beyond the additional interpretability that
sparsity provides, we find further motivation to encourage sparsity of
the NMF estimate when working with networks. For instance, suppose
$(A)_{ij}$=0 for some $i,j$, that is, there is an absense of an edge
between nodes $i$ and $j$. In the low rank approximation there is no
guarentee that $(\hat{A})_{ij}=0$, though we expect it to be near
zero. A straightforward way to force $(\hat{A})_{ij}$ \emph{exactly}
to zero is by anchoring $(U)_{ik} = (V)_{jk} = 0$ for all $k$, and
estimating the remaining elements of $U$ and $V$ by the algorithm
provided below (see \cite{jcgs:mds} for a similar strategy 
for multidimensional scaling).  However, anchoring is not
appropriate with repeated or sequential observations, as an edge can
appear and disappear due to noise.  Keeping in mind the extension to
sequences of networks in the next section, we instead encourage sparsity in
the form of an $l_{1}$ penalty.

The factorized matrices are
obtained through minimizing an objective function that consists of a
goodness of fit component and a roughness penalty
\begin{equation} \label{eqn:sparseNMF}
\min_{U \ge 0, V \ge 0} ||A - U V^T ||_{F}^2 + \lambda_{s} \sum_{k=1}^{K} ||V_{k}||_{1}, 
\end{equation}
where the parameter $\lambda_s \ge 0$. The strength of the penalty is
set by the user to steer the analysis, where a larger penalty
encourages sparser $V$. Adding penalties to NMF is a common strategy, since
they not only improve interpretability, but often improve
numerical stability of the estimation by making the NMF optimization
less under-constrained.
\cite{Berry_Survey, Smooth_NMF,SparseNMF1,SparseNMF2, nnfmGraph1} and
references therein review important penalized NMF models
(see \cite{sPCA, Witten,sblPCA} for similar approaches with SVD).

An advantage of an NMF-based approach is that it is easy to modify for   
particular datasets. For example, a similar $l_{1}$ penalty can be included 
on $U$ 
if the rowspace (typically out-going edges) are of interest.

The estimation algorithm we present is similar to the benchmark
algorithm for NMF, known as `multiplicative
updating' \cite{NNMF1,NNMF2}. The algorithm can be viewed as an
adaptive gradient descent. It is relatively simple to implement, but
can converge slowly due to its linear 
rate \cite{NMF-Optimality}. In practice we find that after a handful
of iterations, the algorithm results in visually meaningful
factorizations.  The estimation algorithm for the penalized NMF in
Eq.~\ref{eqn:sparseNMF} is studied in \cite{SparseNMF1}
and \cite{SparseNMF2}, and the main derivation steps we present next
follow these works.

First, to enforce the non-negativity constraints, we consider the
Lagrangian
\begin{eqnarray}
  L&=&||A - U V^T ||_{F}^2 + \lambda_{s} \sum_{k=1}^{K} ||V_{k}||_{1} \\
   &+& Tr(\Phi U^T) + Tr(\Psi V^T) \nonumber,
\end{eqnarray}
where $\Phi,\Psi$ are Lagrange multipliers. 

To develop a modern gradient descent algorithm, we employ the following 
Karush-Kuhn-Tucker (KKT) optimality conditions, which provide necessary 
conditions for a local minimum \cite{Boyd_Vandenberghe}. The KKT optimality conditions are obtained 
by setting $\frac{\partial L}{\partial U} = \frac{\partial L}{\partial V} = 0$.
\begin{eqnarray}
\Phi &=& -2AV+2UV^T V \\
\Psi &=& -2A^T U + 2VU^T U + 2\lambda_s.
\end{eqnarray} 

Then, the KKT complimentary slackness conditions yield
\begin{eqnarray}
0 &=& (-2AV+2UV^T V)_{ij}(U)_{ij}\\
0 &=& (-2A^T U + 2VU^T U + 2\lambda_s)_{ij}(V)_{ij},
\end{eqnarray}
which, after some algebraic manipulation, lead to the multiplicative 
update rules shown in
Algo. \ref{algo:nmf:static}. The algorithm has some notable
theoretical properties. Specifically, each iteration of the algorithm
will produce estimates that reduce the objective function value, e.g.,
the estimates improve at each iteration. Minor modifications provided
in \cite{nmf:mult:lin} can be employed to guarantee convergence to a
stationary point.

\begin{algorithm} 
\begin{algorithmic}[1]
\STATE Set constant $\lambda_{s}$
\STATE Initialize $\{U,V\}$ as dense, positive random matrices
\REPEAT
\STATE Set
\begin{equation}
(U)_{ij} \leftarrow (U)_{ij} \frac{(A V)_{ij}}{(UV^T V)_{ij}} \nonumber
\end{equation}
\STATE Set
\begin{equation}
(V)_{ij} \leftarrow (V)_{ij} \frac{(A^T U)_{ij}} {(V U^T U)_{ij}+\lambda_s} \nonumber
\end{equation}
\UNTIL{Convergence}
\end{algorithmic}
\caption{Sparse NMF}
\label{algo:nmf:static}
\end{algorithm} 

Lastly, we note that when the observed graph is undirected, 
due to symmetry of the adjacency matrix the factorization can be 
written as 
\begin{equation}
A \approx U\Lambda U^T,
\end{equation}
where $\Lambda$ is a non-negative diagonal matrix. This is the underlying model 
investigated in Facetnet \cite{facetnet}, with additional constraints on $U$ to satisfy 
an underlying probabilistic interpretation. The objective function considered 
in \cite{facetnet} was also based on relative entropy or KL-divergence. 
We find that such symmetric NMF models are  
far more sensitive to additional constraints than its general
counterpart, especially when dealing with sequences of networks as in
the next section. Symmetric NMF has less flexibility, since additional
constraints strongly influence the reconstruction accuracy of the
estimation. On the other hand, without imposing symmetry, as $V$
changes, $U$ compensates (and vice versa) in order for the final
product to reproduce the data as best as possible. Thus, for tasks of
visualization of node evolution and community extraction in dynamic
networks, we do not impose symmetry on the factorization.

\subsection{Illustrative Examples}
\subsubsection{Community Discovery on a Toy Example}
\begin{figure}
\centerline{
\includegraphics[width=1\columnwidth,trim=3cm 3cm 2.2cm 2cm, clip=true]{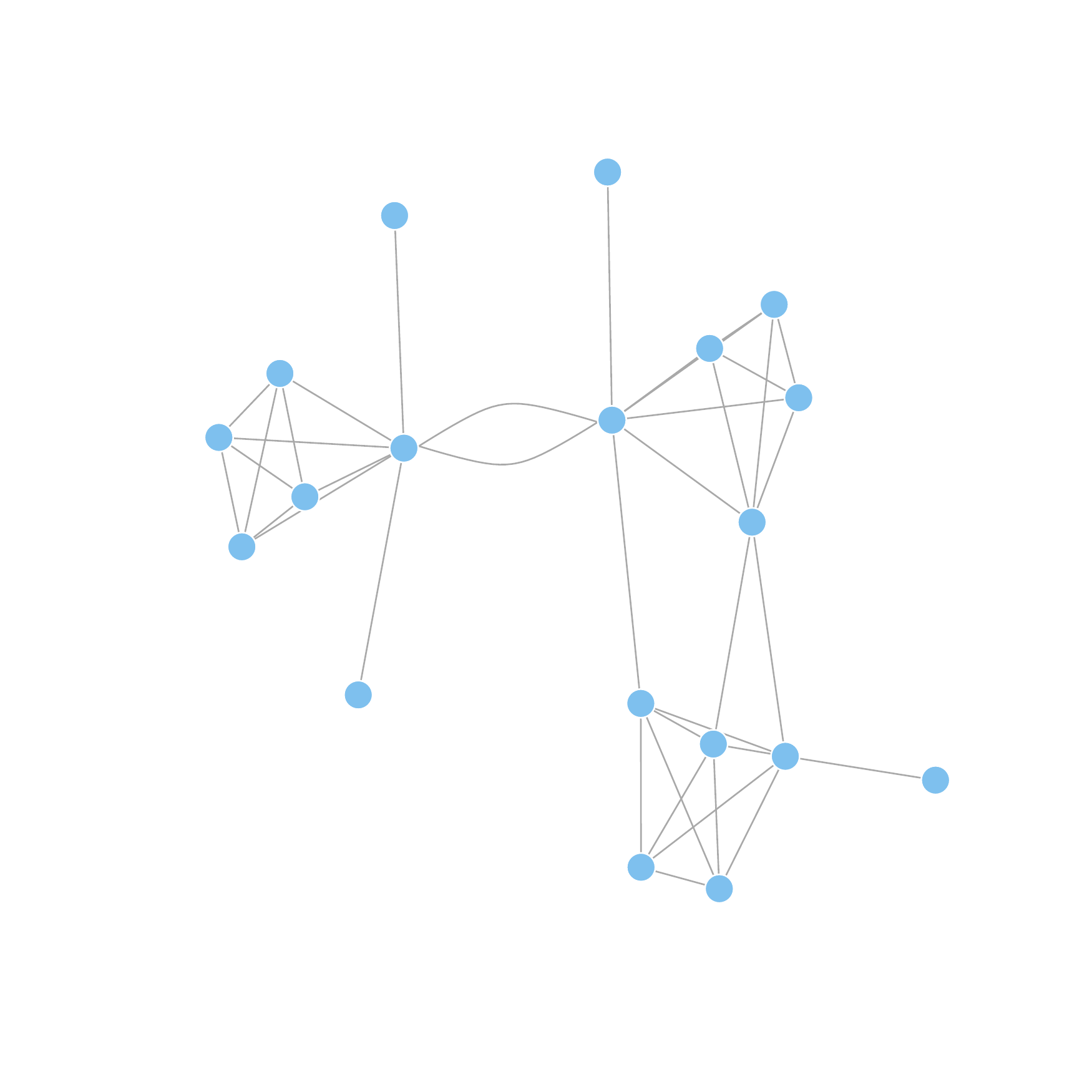}
}
\caption{An undirected network with 19 nodes.}
\label{fig:static:toy}
\end{figure}

We compare the following methods on a toy example shown in Fig.~\ref{fig:static:toy}.
\begin{enumerate} 
	\item Leading eigenvector (modularity) based community discovery \cite{newman-modularity-matrix}
	\item Spectral clustering \cite{1227.62042}
	\item Clique percolation \cite{2005Natur.435..814P} for overlapping community discovery
	\item Classical NMF (Eq.~\ref{obj:simple})
	\item Sparse NMF (Eq.~\ref{eqn:sparseNMF})
\end{enumerate}

\begin{figure*}
\begin{tabular}{cccc}
Leading eigenvector & Spectral clustering & Clique percolation & Classical NMF\\
\includegraphics[width=0.5\columnwidth,trim=3cm 3cm 2.2cm 2cm, clip=true]{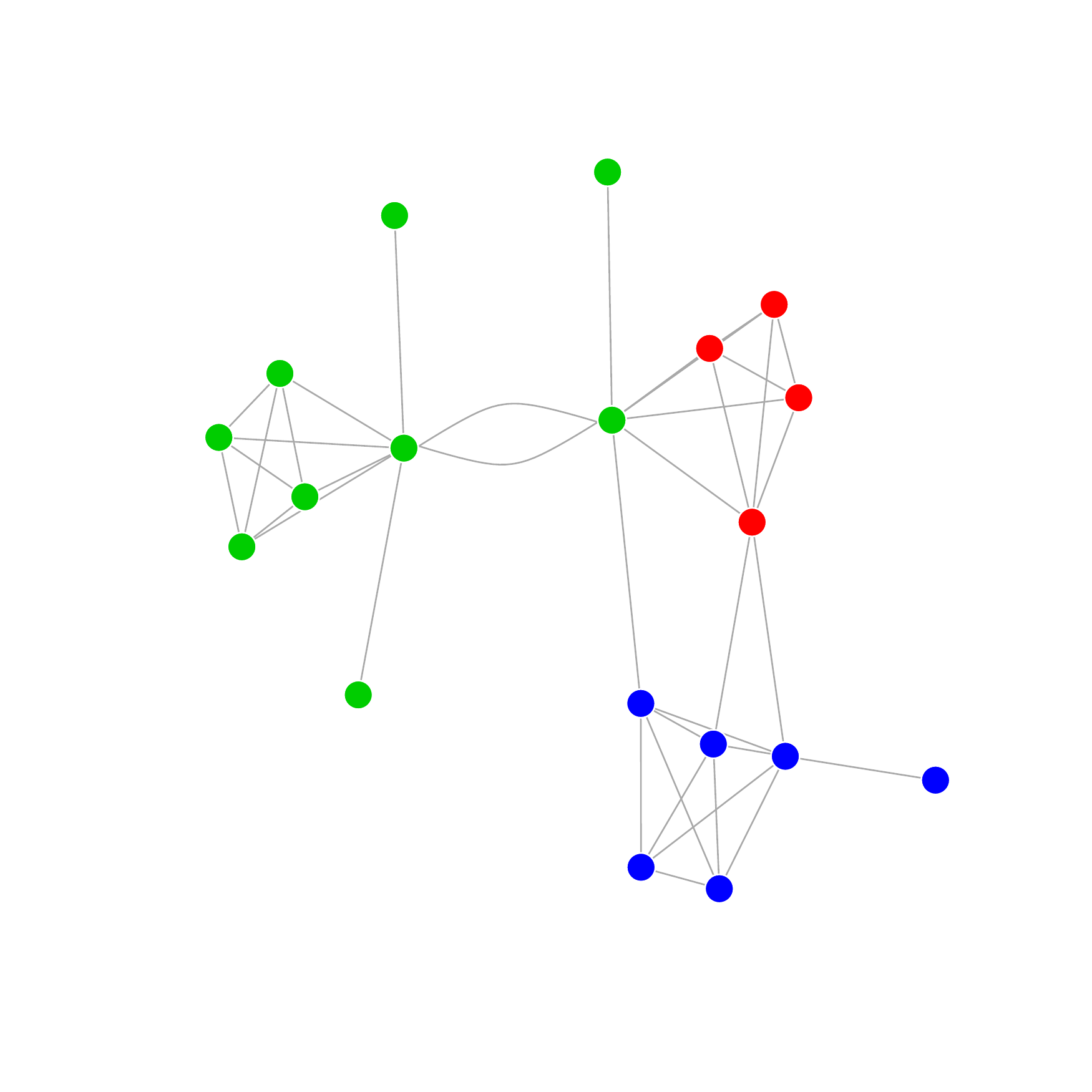} &
\includegraphics[width=0.5\columnwidth,trim=3cm 3cm 2.2cm 2cm, clip=true]{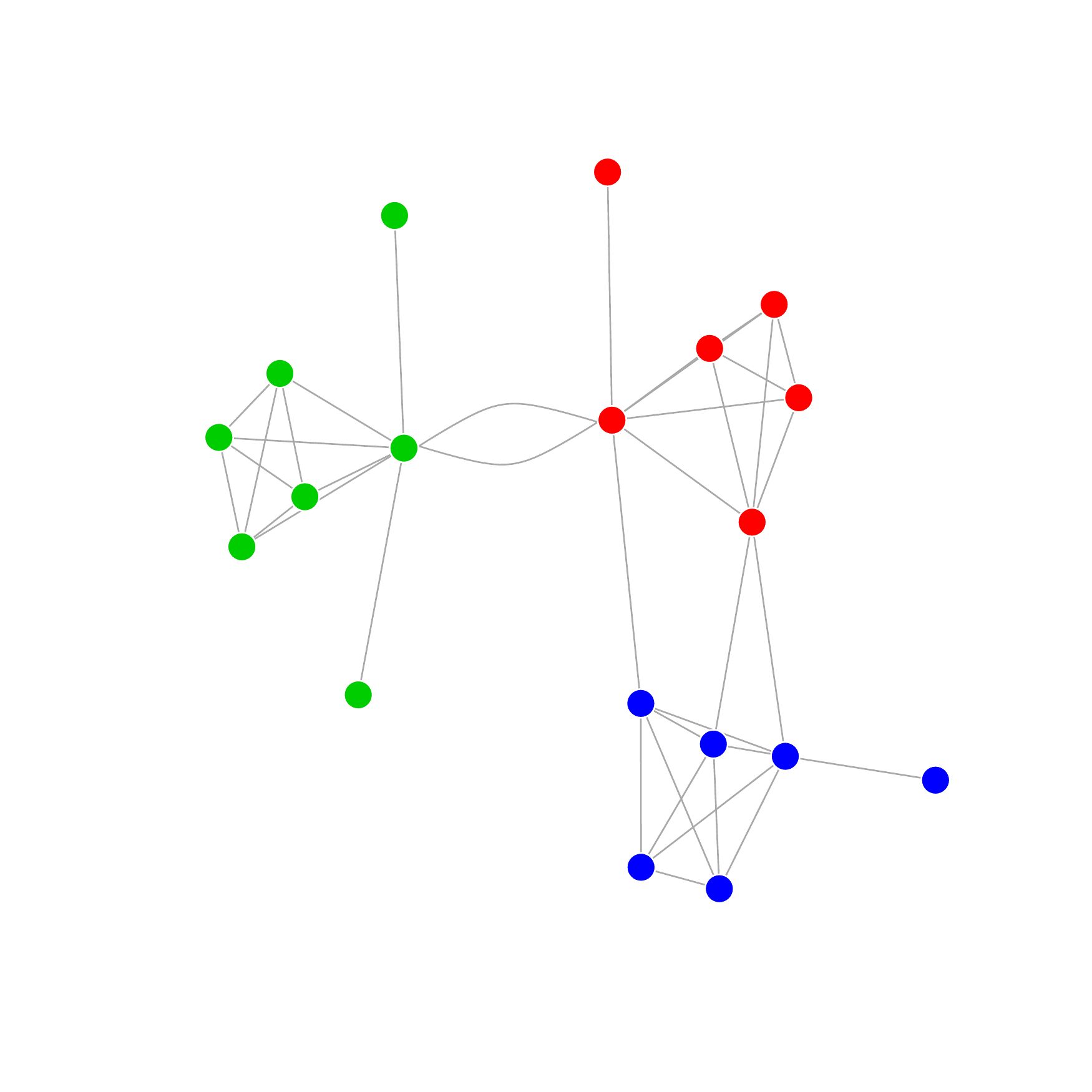} &
\includegraphics[width=0.5\columnwidth,trim=3cm 3cm 2.2cm 2cm, clip=true]{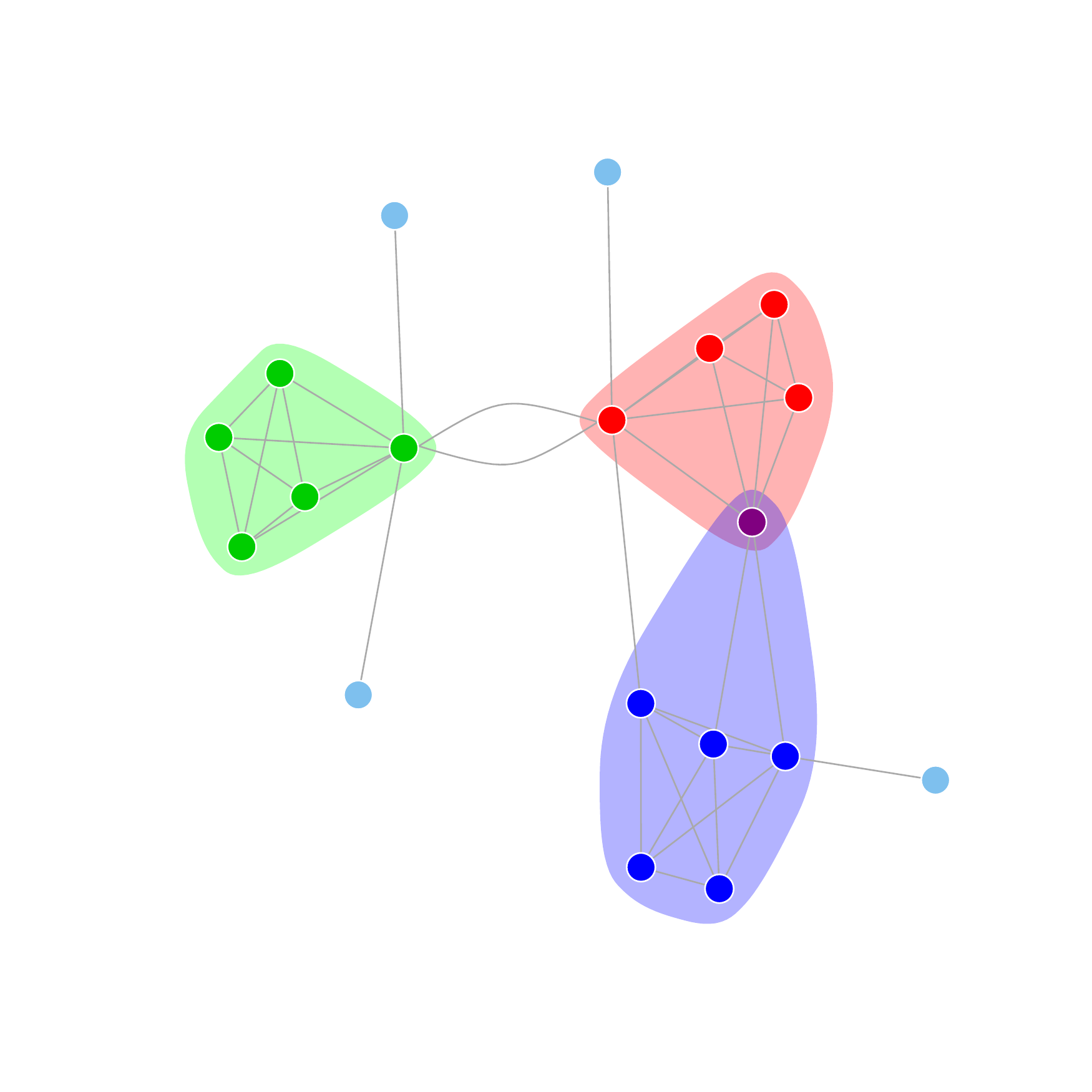} & 
\includegraphics[width=0.5\columnwidth,trim=3cm 3cm 2.2cm 2cm, clip=true]{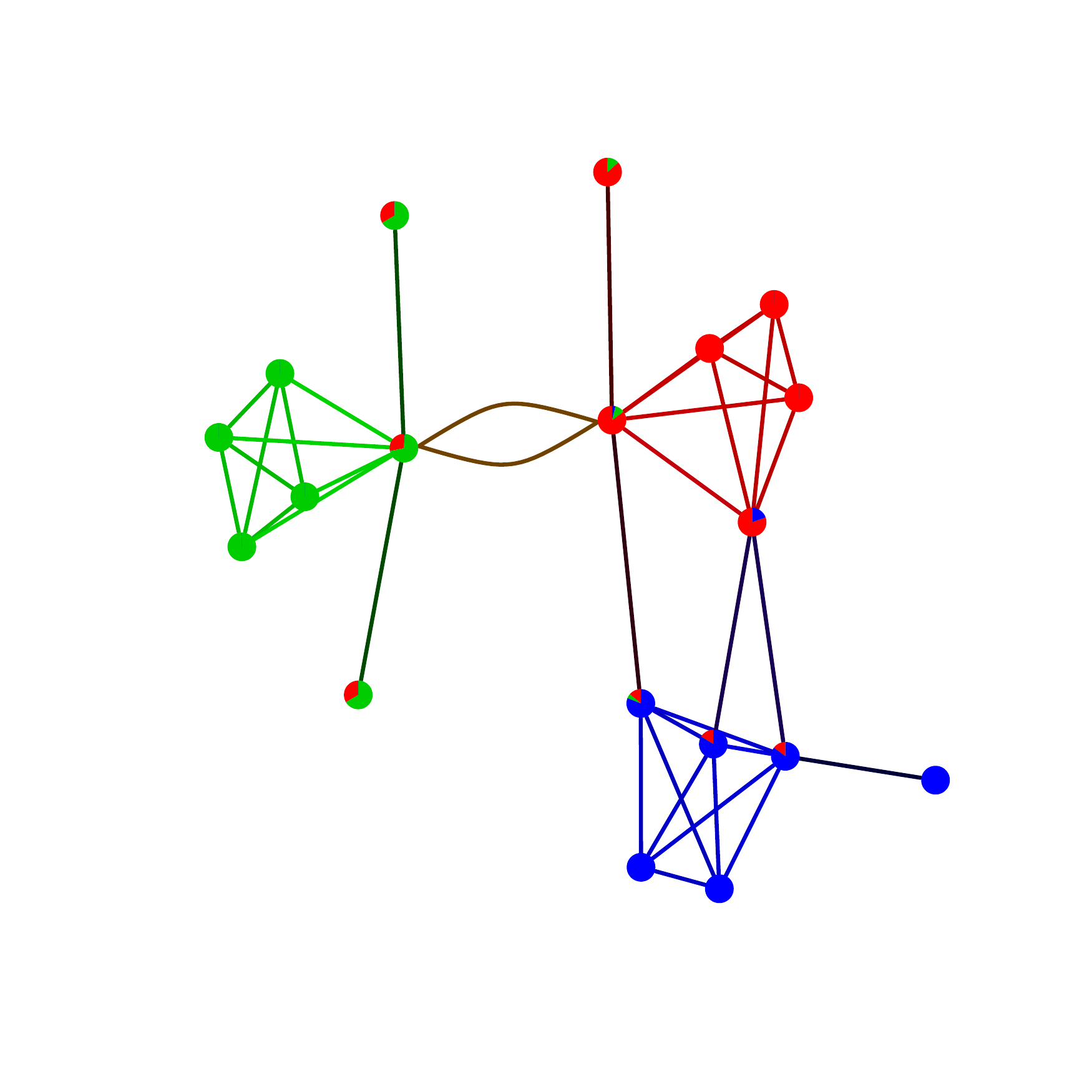}
\end{tabular}
\caption{(Color online) Results using alternative community discovery methods.}
\label{fig:static:competitors}
\end{figure*}

\begin{figure}
\includegraphics[width=1\columnwidth,trim=3cm 3cm 2.2cm 2cm, clip=true]{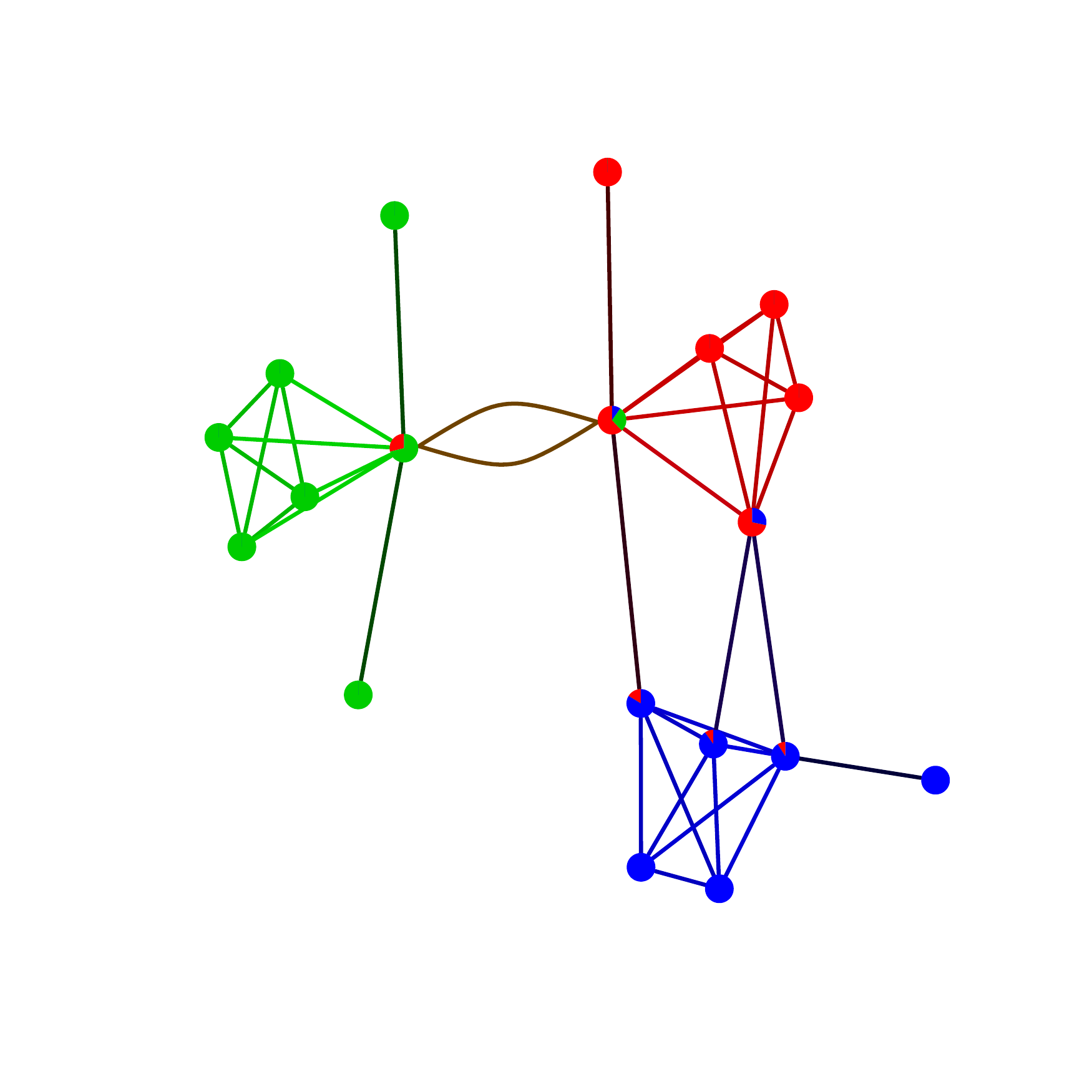}
\caption{(Color online) Results from applying sparse NMF (Algo.~\ref{algo:nmf:static}) with $\lambda_{s} = 5$. Nodes and edges are colored to denote the relative contribution of each community.}
\label{fig:static:sparseNMF}
\end{figure}

The results of the alternative methodologies are provided in Fig.~\ref{fig:static:competitors}, where we see that even on this toy example, there is disagreement in the recovered community structure. The leading eigenvector solution differs slightly from that of spectral clustering. Taken together, one may suspect a soft partitioning would result in overlap between the green and red communities. Yet, clique percolation finds overlap between the blue and red communities. Classical NMF finds overlap between all three communities, quantifies the amount of overlap (denoted by the pie chart on each node), and decomposes each edge by community (colored as a mixture of red, green and blue). 
Fig.~\ref{fig:static:sparseNMF} shows that sparse NMF finds a cleaner structure compared to classical NMF. In particular, the sparse NMF solution has less overlap (mixing) between the three groups, while still quantifying community contribution to nodes and edges.

\begin{figure}
\begin{tabular}{ccc}
 & Star Network & Ring Network \\
 & \includegraphics[width=0.25\columnwidth]{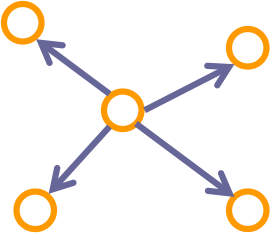} & \includegraphics[width=0.3\columnwidth]{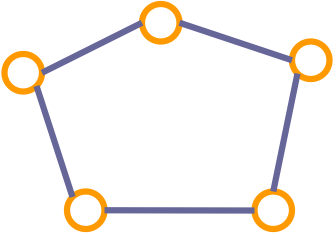} \\
\begin{tabular}{c}
Adjacency\\
Matrix
\end{tabular} &
   	    $\left[ \begin{array}{ccccc}
		0 & 1 & 1 & 1 & 1\\
		0 & 0 & 0 & 0 & 0\\
		0 & 0 & 0 & 0 & 0\\
		0 & 0 & 0 & 0 & 0\\
		0 & 0 & 0 & 0 & 0
		 \end{array} \right]$
& 
   	    $\left[ \begin{array}{ccccc}
		0 & 1 & 0 & 0 & 1\\
		1 & 0 & 1 & 0 & 0\\
		0 & 1 & 0 & 1 & 0\\
		0 & 0 & 1 & 0 & 1\\
		1 & 0 & 0 & 1 & 0
		 \end{array} \right]$ \\
\begin{tabular}{c}
NMF \\
$U,V$
\end{tabular} & 
$\left[ \begin{array}{r}
		1.00 \\
		0.00 \\
		0.00 \\
		0.00 \\
		0.00 
		 \end{array} \right]
\left[ \begin{array}{r}
		0.00 \\
		1.00 \\
		1.00 \\
		1.00 \\
		1.00 
		 \end{array} \right]$		 
&  
$\left[ \begin{array}{r}
		0.89 \\
		0.89 \\
		0.89 \\
		0.89 \\
		0.89 
		 \end{array} \right]
\left[ \begin{array}{r}
		0.45 \\
		0.45 \\
		0.45 \\
		0.45 \\
		0.45 
		 \end{array} \right]$\\
\begin{tabular}{c}
Hub, Authority
\end{tabular} &
$\left[ \begin{array}{r}
		1.00 \\
		0.00 \\
		0.00 \\
		0.00 \\
		0.00 
		 \end{array} \right]
\left[ \begin{array}{r}
		0.00 \\ 
		0.50  \\
		0.50  \\
		0.50  \\
		0.50  
		 \end{array} \right]$		 
& 
$\left[ \begin{array}{r}
		0.45 \\
		0.45 \\
		0.45 \\
		0.45 \\
		0.45 
		 \end{array} \right]
\left[ \begin{array}{r}
		0.45 \\
		0.45 \\
		0.45 \\
		0.45 \\
		0.45 
		 \end{array} \right]$		 
\end{tabular}
\caption{Rank 1 NMF without penalization and Kleinberg's authority/hub scores \cite{kleinberg-authority}.}
\label{fig:illustrative}
\end{figure}

\subsubsection{Rank One Factorizations}
We show in our experiments (Section \ref{sec:experiments}) that a sequence of rank one 
matrix factorizations can be the basis for informative displays of 
time-varying node importance to connectivity. 
To provide some intuition as to why such a rank one factorization is informative, 
consider Fig. \ref{fig:illustrative}, which shows graph structures, corresponding NMFs,  
and Kleinberg's authority and hub scores \cite{kleinberg-authority}. Authority and hub scores 
are computed by the leading eigenvector of $A^{T}A$ and $AA^{T}$, respectively. 
Subject to rescaling of the 
NMF estimates, the results are identical. In fact, by the Perron-Frobenius theorem (see Chapter 8 of \cite{Meyer:2000:MAA:343374}), the rank one  
NMF solution is always a rescaled version of authority and hub scores. 
This provides a natural interpretation for the rank one NMF. For instance, the $U$ vector on 
the Star Network highlights the hub node.  
The $V$ vector show that all peripheral nodes are equal
in terms of their authority (incoming connections), and that the central node has
no incoming connections. NMF vectors of the Ring Network show each node with an
equal score for incoming (authority) and outgoing (hub) connectivity. The fact that $U$
contains larger elements than $V$ is arbitrary. However, the
assignment of equal values within $U$ and $V$ shows each node is
equally important to interconnectivity.

\section{Model for Dynamic Networks} \label{sec:overview}
Given a time series of networks $\{\mathcal{G}_t = (V_t,
E_t)\}_{t=1}^{T}$ with corresponding adjacency matrices
$\{A_t\}_{t=1}^{T}$, the goal is to produce a sequence of low rank
matrix factorizations $\{U_{t},V_{t}\}_{t=1}^{T}$.

To extend the factorization from the previous section 
to the temporal setting, we impose a
smoothness constraint on the basis $U_{t}$. This constraint 
forces new community structure to be similar to previous time points. 
Since individual node
time-series given by $U_{t}$ are visually smooth, 
time plots for each node become informative 
and provide an alternative to graph drawings for visualizing node dynamics. 
Moreover, time plots 
are static displays, which facilitate detailed analysis and avoid
difficulties associated with animated layouts when given a large
number of time points or nodes.

The objective function becomes
\begin{eqnarray}
\label{obj:full}
\min_{\{U_{t} \ge 0, V_{t} \ge 0\}_{t=1}^{T}}&& \sum_{t=1}^{T} ||A_t - U_{t} V_{t}^T ||_{F}^2 \\
&&{+}\: \lambda_{t} \sum_{t=1}^{T}\sum_{\tilde{t}=t-\frac{W}{2}}^{t+\frac{W}{2}} ||U_{t}-U_{\tilde{t}} ||_{F}^2  \nonumber \\
&&{+}\: \lambda_{s} \sum_{t=1}^{T} \sum_{k=1}^{K} ||V_{t,k}||_{1}, \nonumber
\end{eqnarray}
where $W$ is a small integer representing a time window. The
parameters $\lambda_t,\lambda_s \text{ and } W$ are 
set by the user to steer the analysis. 

The interpretations of $U_{t}, V_{t}$ extend naturally from the
previous section, so that, for instance, $\sum_{k}(V_t)_{kj}$ measures
the importance of node $j$ (typically corresponding to incoming
edges), and $(U_t)_{ik} (V_t)_{jk} / \sum_{k=1}^{K}
(U_t)_{ik}(V_t)_{jk}$ to measure the relative contribution of each
community to each $i,j$ edge. In principle, the edge decomposition can
be used to assign nodes to communities as discussed in the last
section.  However, this approach can be unsatisfactory due to unstable
community assignments. As alternative method is to assign communities
in terms of $U_{t}$, which ensures the stability of the community
structure through time. Specifically, measuring the contribution of
node $i$ to each community with the relative magnitude of the $i$th
element of each dimension of $U_{t}$, e.g., $(U_t)_{ik}
/ \sum_{k=1}^{K} (U_t)_{ik}$.

We can follow similar steps as in the last section to derive a 
gradient descent estimation algorithm. 
First, to enforce the non-negativity constraints, we consider the
Lagrangian
\begin{eqnarray}
  L&=&\sum_{t=1}^{T} ||A_t - U_{t} V_{t}^T ||_{F}^2 \\
  &&{+}\: \lambda_{t} \sum_{t=1}^{T}\sum_{\tilde{t}=t-\frac{W}{2}}^{t+\frac{W}{2}}
  ||U_{t}-U_{\tilde{t}} ||_{F}^2 \nonumber \\
  &&{+}\: \lambda_{s} \sum_{t=1}^{T} \sum_{i=1}^{n} \sum_{j=1}^{K} |V_t (i,j)| \nonumber \\
  &&{+}\: \sum_{t=1}^{T} Tr(\Phi_{t} U_{t}^T) + \sum_{t=1}^{T}Tr(\Psi_{t} V_{t}^T) \nonumber ,
\end{eqnarray}
where $\Phi_{t},\Psi_{t}$ are Lagrange multipliers. 

The following KKT optimality conditions are obtained by setting
$\frac{\partial L}{\partial U_{t}} = \frac{\partial L}{\partial V_{t}}
= 0$.
\begin{eqnarray}
\Phi_{t}&=&-2A_{t}V_{t}+2U_{t}V_{t}^T V_{t} - 2 \lambda_{t}\sum_{\tilde{t}=t-\frac{W}{2}}^{t-1} U_{\tilde{t}}\\
&&{-}\: 2 \lambda_{t}\sum_{\tilde{t}=t+1}^{t+\frac{W}{2}} U_{\tilde{t}}+ 2W\lambda_{t} U_{t}\nonumber \\
\Psi_{t} &=& -2A_{t}^T U_{t} + 2V_{t}U_{t}^T U_{t} + 2\lambda_s.
\end{eqnarray} 

Then, the KKT complimentary slackness conditions yield
\begin{eqnarray}
0&=&(-2A_{t}V_{t}+2U_{t}V_{t}^T V_{t} - 2 \lambda_{t}\sum_{\tilde{t}=t-\frac{W}{2}}^{t-1} U_{\tilde{t}})_{ij}(U_{t})_{ij}\\
&&{+}\: (-2 \lambda_{t}\sum_{\tilde{t}=t+1}^{t+\frac{W}{2}} U_{\tilde{t}}+ 2W\lambda_{t} U_{t})_{ij}(U_{t})_{ij}\nonumber \\
0 &=& (-2A_{t}^T U_{t} + 2V_{t}U_{t}^T U_{t} + 2\lambda_s)_{ij}(V_{t})_{ij},
\end{eqnarray} 
which after some algebra leads to the algorithm provided in Algo.~\ref{algo:nmf:dynamic}. The theoretical 
properties are also the same as in the previous section. Most notably, the estimates of $U_{t}$ and $V_{t}$ will improve at each 
iteration with respect to Eq.~\ref{obj:full}.

\begin{algorithm} 
\begin{algorithmic}[1]
\STATE Set constants $\lambda_{t}, \lambda_{s}, W$.
\STATE Initialize $\{U_t\},\{V_t\}$ as dense, positive random matrices.
\REPEAT
\FOR {t=1..T}
\STATE Set\\
$(U_t)_{ij} \leftarrow (U_t)_{ij} \frac{(A_t V_t
+ \lambda_{t}\sum_{\tilde{t}=t-\frac{W}{2}}^{t-1}
U_{\tilde{t}}+\lambda_{t}\sum_{\tilde{t}=t+1}^{t+\frac{W}{2}}
U_{\tilde{t}})_{ij}}{ (U_{t}V_{t}^T V_{t} + W\lambda_{t}
U_{t})_{ij}}$.
\STATE Set
\begin{equation}
(V_t)_{ij} \leftarrow (V_t)_{ij} \frac{(A_{t}^T U_t)_{ij}} {(V_{t}
U_{t}^T U_t)_{ij}+\lambda_s}. \nonumber
\end{equation}
\ENDFOR
\UNTIL{Convergence}
\end{algorithmic}
\caption{NMF with temporal and sparsity penalties}
\label{algo:nmf:dynamic}
\end{algorithm} 

\subsection{Parameter Selection}
We briefly discuss the important practical matter of choosing $K$, the
inner rank of the matrix factorization.

For the goal of clustering, the rank should be equal to the number of
underlying groups. The rank can be ascertained by examining the accuracy of
the reconstruction as a function of rank. However, this tends to rely
on subjective judgments and overfit the given data. Cross validation
based approaches are theoretically preferable and follow the same
intuition.

The idea behind cross validation is to use random subsets of the data
from each data slice to fit the model, and another subset from each
data slice to assess accuracy. Different values of $K$ are then cycled
over and the one that corresponds to the lowest test error is chosen.

Due to the data structure, we employ two-dimensional cross
validation. Two-dimensional refers to the selection
of \emph{submatrices} for our training and test data. Special care is
taken to ensure that the same rows and columns are held out of every
data slice, and the dimensions of the training and test sets are
identical.

The hold out pattern divides the rows into $k$ groups, the columns
into $l$ groups, then uses the corresponding $kl$ submatrices to fit
and test the model. In each submatrix, the given row and column group
identifies a held out submatrix that is used as test data, while the
remaining cells are used for training. The algorithm is shown in
Algo.~\ref{algo:CV}. The notation in the algorithm uses
$\mathcal{I}_{l}$ and $\mathcal{I}_{J}$ as index sets to identify
submatrices in the each data matrix.

\begin{algorithm}
\begin{algorithmic}[1]
\STATE Form row holdout set: $\mathcal{I}_{l} \subset \{1,..,n\}$
\STATE Form column holdout set: $\mathcal{I}_{J} = \subset \{1,..,n\}$
\STATE Set 
\begin{equation*}
	(\tilde{U}_{t}, \tilde{V}_{t}) = \argmin_{U_{t},V_{t} \ge 0} \sum_{t}||(A_{t})_{-\mathcal{I}_{l}, -\mathcal{I}_{J}} - U_{t}V_{t}^T ||_{F}^2
\end{equation*}
\STATE Set 
\begin{equation*}
	\breve{U_{t}} = \argmin_{U_{t} \ge 0} \sum_{t}||(A_{t})_{\mathcal{I}_{l}, -\mathcal{I}_{J}} - U_{t}\tilde{V}_{t}^T ||_{F}^2
\end{equation*}
\STATE Set 
\begin{equation*}
	\breve{V}_{t} = \argmin_{V_{t} \ge 0}  \sum_{t}||(A_{t})_{-\mathcal{I}_{l}, \mathcal{I}_{J}} - \tilde{U}_{t}V_{t}^T ||_{F}^2
\end{equation*}
\STATE Set 
\begin{equation*}
	\hat{(A_{t})}_{\mathcal{I}_{l}, \mathcal{I}_{J}} = \breve{U}_{t}\breve{V}_{t}^T
\end{equation*}
\STATE Compute Test error 
\begin{equation*}
	\text{Test Error}= \sum_{t}||(A_{t})_{\mathcal{I}_{l}, \mathcal{I}_{J}} - \hat{(A_{t})}_{\mathcal{I}_{l}, \mathcal{I}_{J}}||_{F}^2
\end{equation*}
\end{algorithmic}
\caption{Cross-validation for choosing the number of communities (rank)}
\label{algo:CV}
\end{algorithm}

\begin{figure}
\centerline{
\includegraphics[width=1\columnwidth]{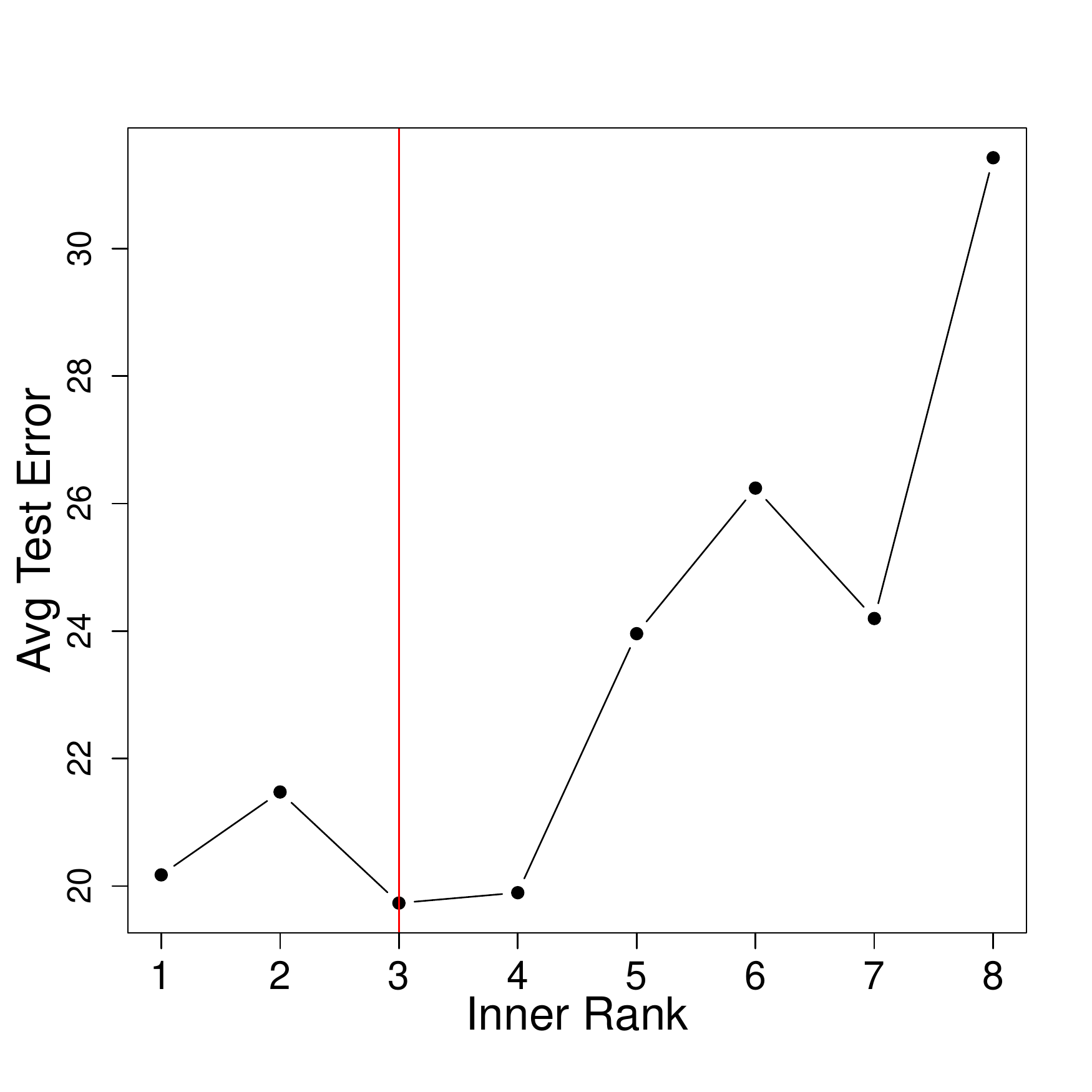}
}
\caption{Cross validation indicates 3 communities (rank 3) features the lowest average test error for the toy example.}
\label{fig:CV:static}
\end{figure}

We then cycle over different values of $K$ to choose the one that
minimizes average test error. 
Fig.~\ref{fig:CV:static} shows that this procedure 
correctly identifies 3 communities for the toy example. Consistency results are developed
in \cite{owen_CV} to provide theoretical foundations for this 
approach. 

In principle the cross validation procedure can be used to select the
penalties $\lambda_{t},\lambda_{s}$ and the time window $W$. However,
considering the scale of many modern network datasets, this would
require too much computing time. Instead we typically choose the
penalties by hand to emphasize readability and interpretability of the
results, keeping in mind that if either penalty is set too large then
the estimation results in degenerate solutions.  For instance, the
algorithm suffers from numerical instabilities when $\lambda_s$ is too
large, since all $V_{t}$ elements are zero.  If $\lambda_t$ is set to
an extremely large number, then $U_{t}$ will be approximately constant
for all time periods, so the effective model is $A_{t}\approx
UV_{t}^{T}$, e.g., the community structure is fixed for all
observations.

The parameter, $W$, controls the
number of neighboring time steps to locally average. Larger values of
$W$ mean that the model has more memory so it incorporates more time
points for estimation. This risks missing sharper changes in the data
and only detecting the most persistent patterns. On the other hand,
small values of $W$ make the fitting more sensitive to sharp changes,
but increase short term fluctuations due to smaller number of
observations. We set $W = 2$ (looking one time period ahead and
before) for all presented experiments. Larger values could be used in
very noisy settings to further smooth results.

\section{Experiments} \label{sec:experiments}
In this section we test the model on both synthetic and real-world
examples. The synthetic networks allow us to validate the model's
ability to highlight known community structure and node evolution,
while the real examples exhibit the model's performance under
practical conditions.

\subsection{Synthetic Networks}
\begin{figure}
\centerline{
\scalebox{0.6}{\includegraphics[trim=3cm 3cm 2cm 2cm, clip=true]{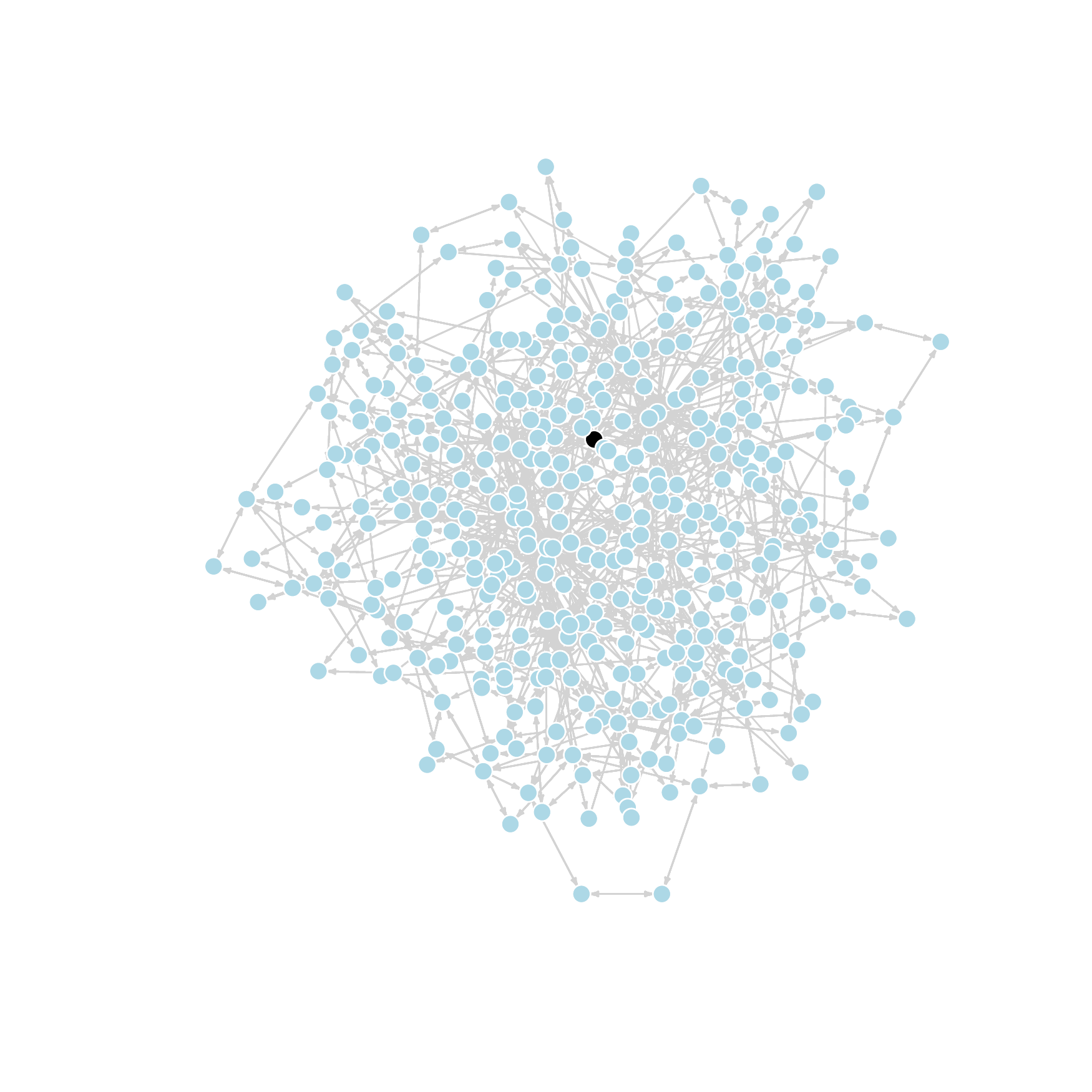}}
}
\caption{The cell phone network from a day using a force directed 
  layout algorithm in igraph. Node 200 is colored black.}
\label{fig:vast-raw}
\end{figure}

\begin{figure*}
\centerline{
\scalebox{0.35}{\includegraphics{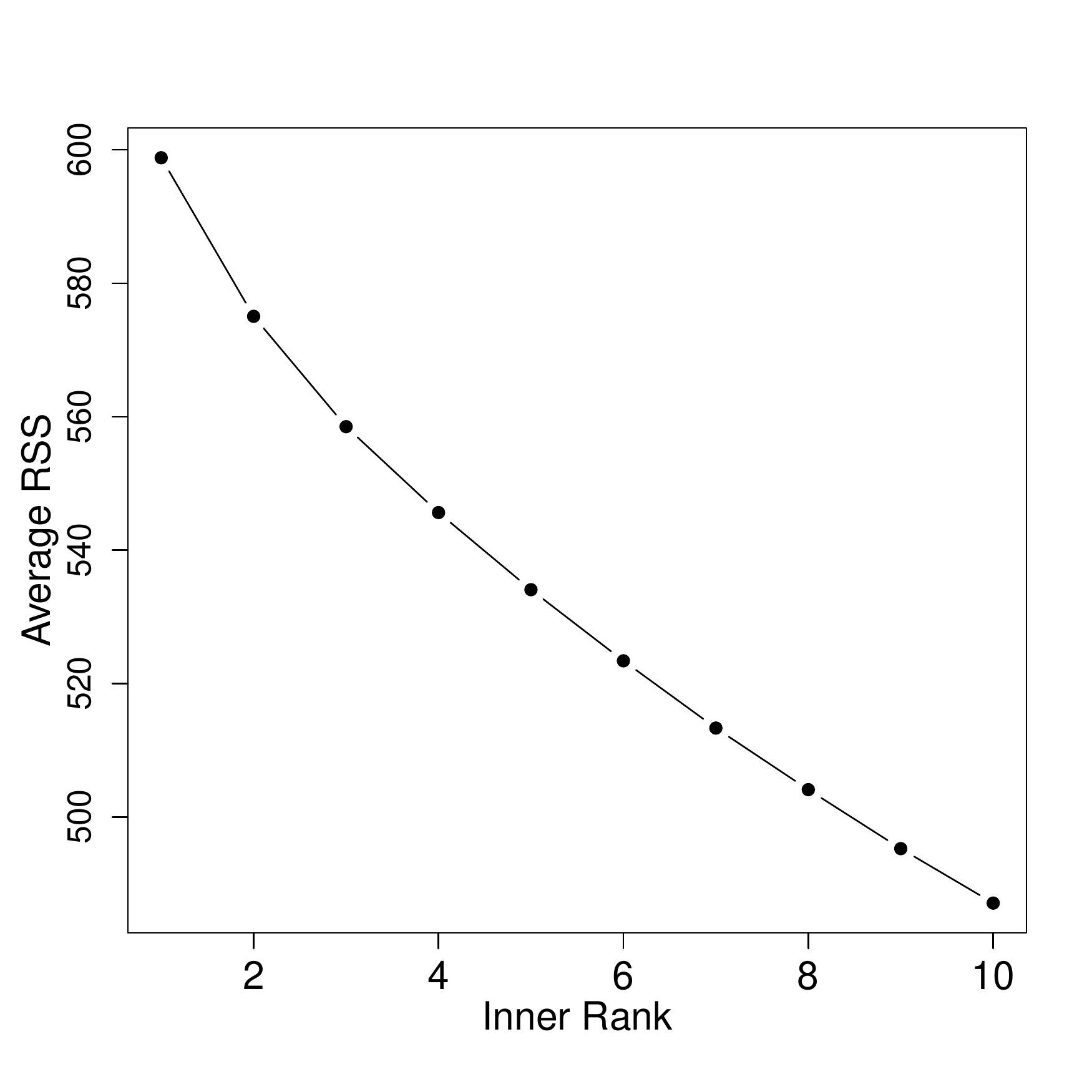}}
\scalebox{0.35}{\includegraphics{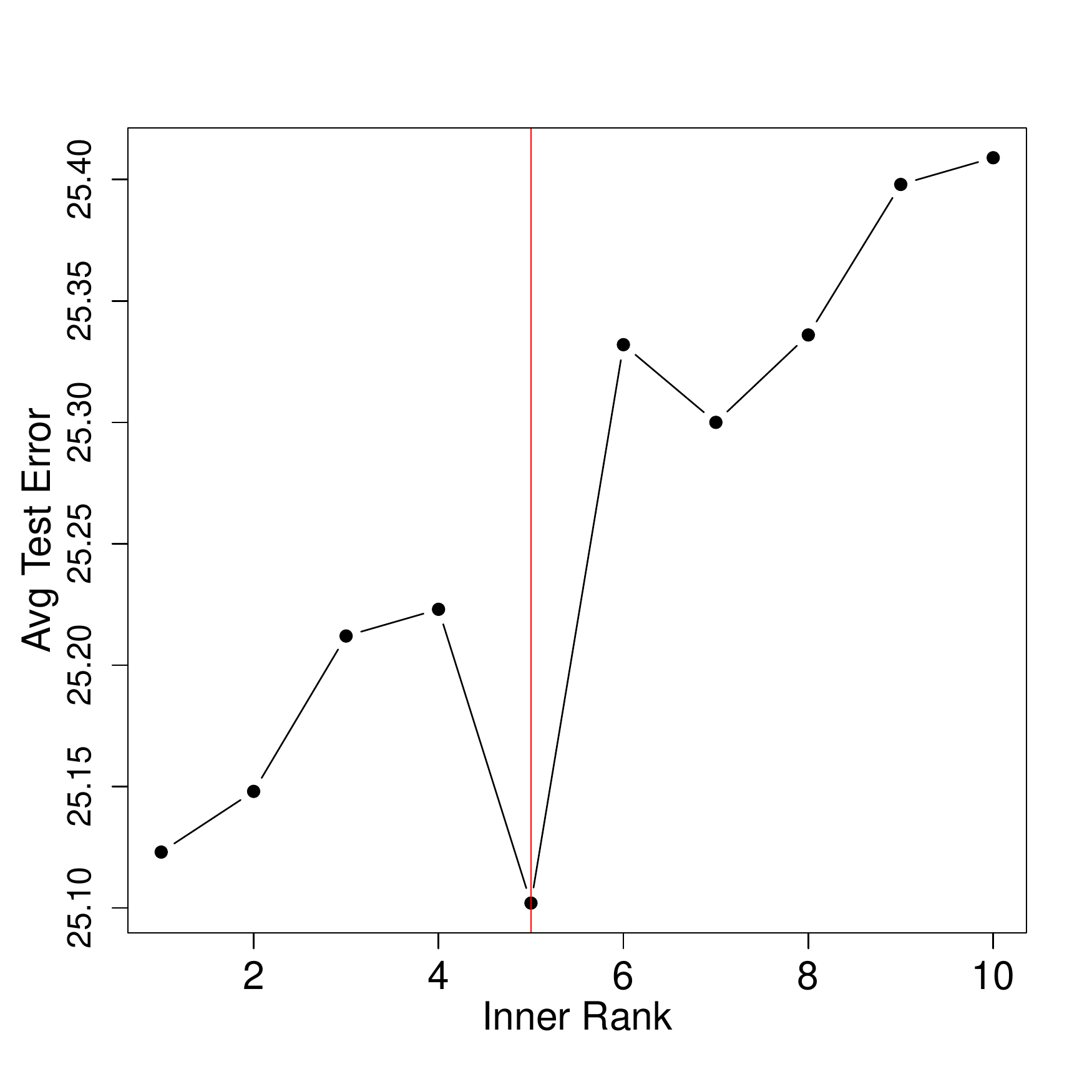}}
}
\caption{Choosing $K$ for the Catalano communications network. The
  left panel shows the average residual sum of squares, and the right
  panel shows the average test error obtained via cross validation ($5\times5$ fold)
  for different the approximation ranks. Cross validation indicates that 5 
  communities is optimal.}
\label{fig:vast}
\end{figure*}

\begin{figure*}
\centerline{
\begin{tabular}{lccc}
 & Day 5 & Day 6 & Day 7 \\
\raisebox{3cm}{Raw} & \includegraphics[width=0.6\columnwidth, trim=3cm 3cm 2.75cm 2.75cm, clip=true]{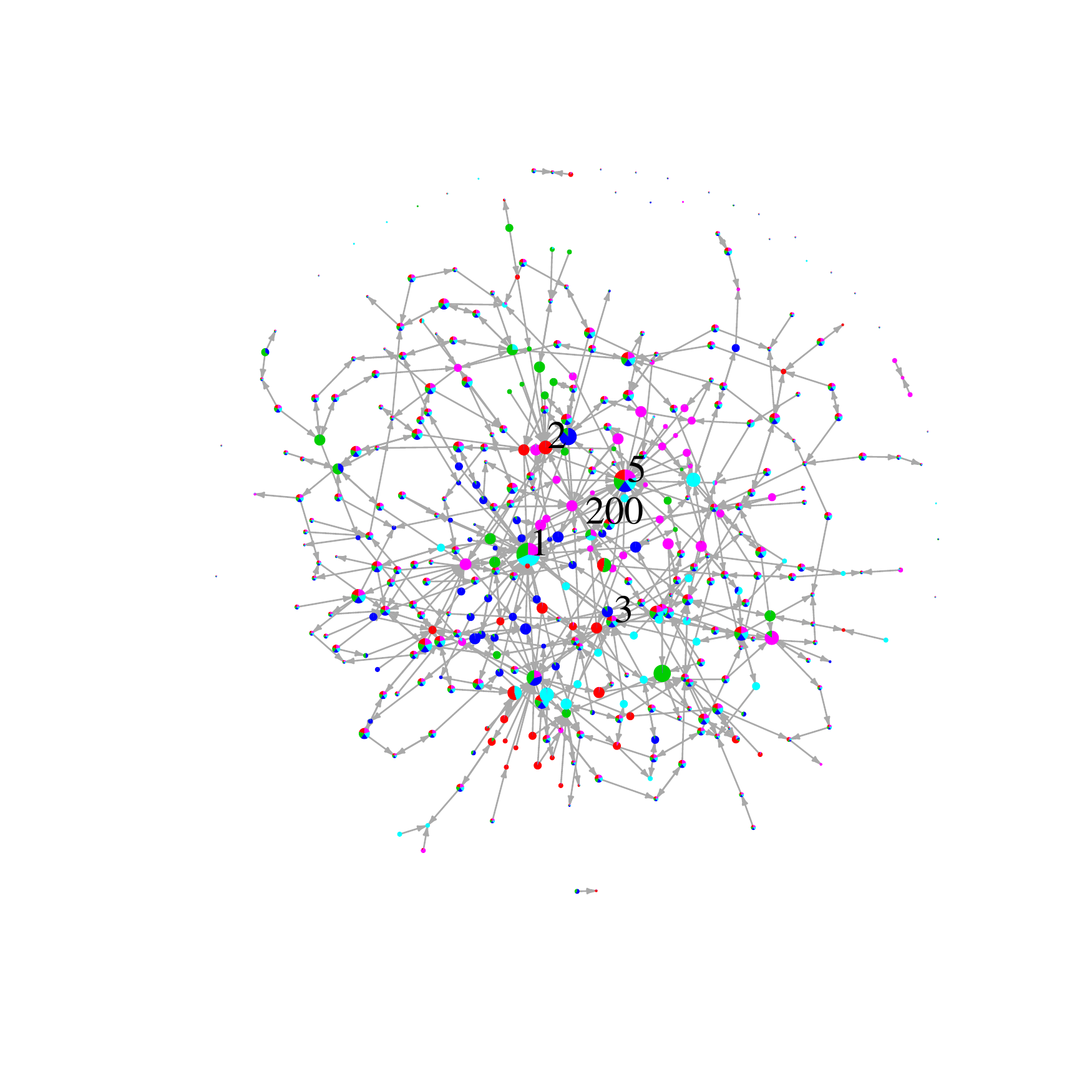}& 
\includegraphics[width=0.6\columnwidth, trim=3cm 3cm 2.75cm 2.75cm, clip=true]{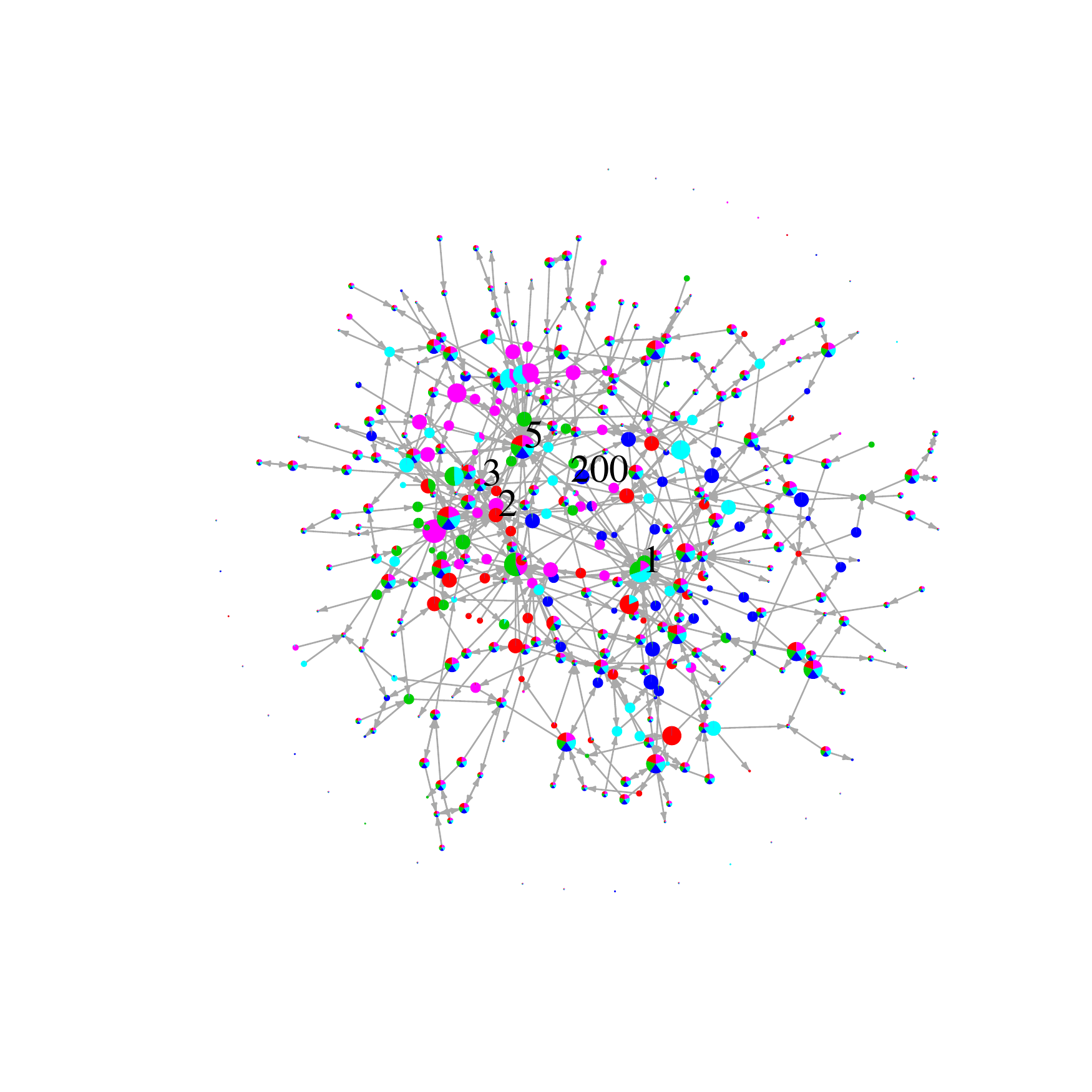}& 
\includegraphics[width=0.6\columnwidth, trim=3cm 3cm 2.75cm 2.75cm, clip=true]{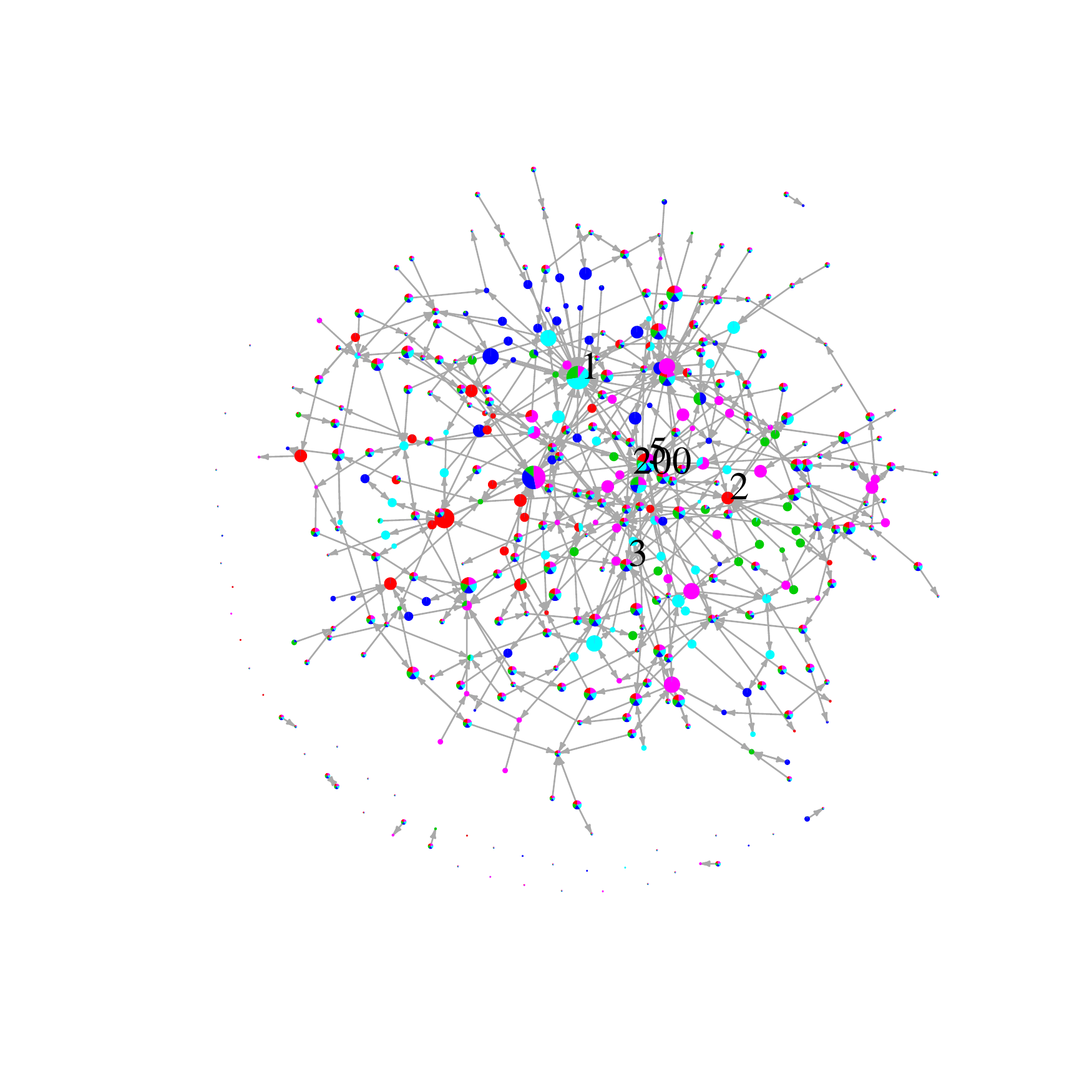}\\
\raisebox{3cm}{Fitted} & \includegraphics[width=0.6\columnwidth, trim=3cm 3cm 2.75cm 2.75cm, clip=true]{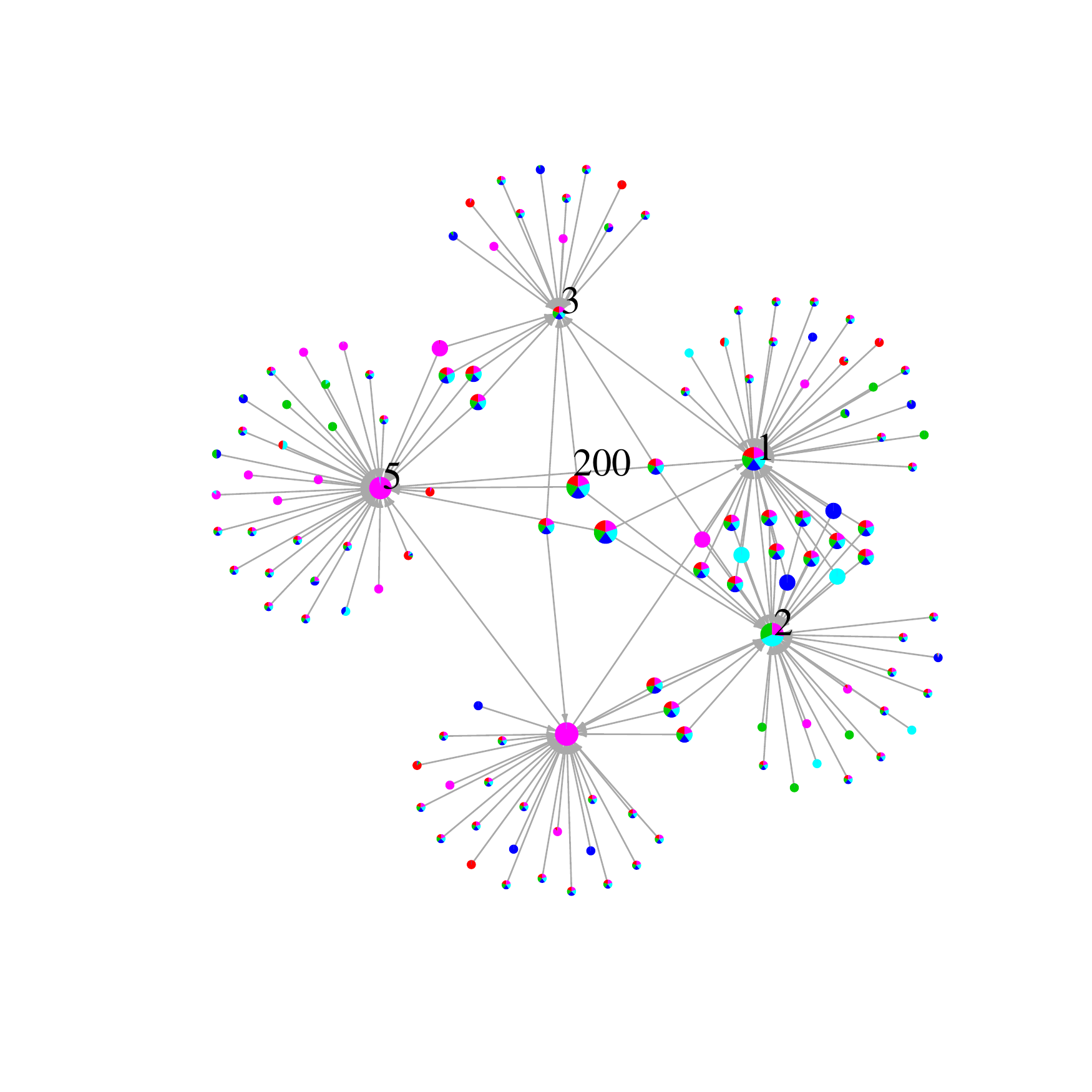}& 
\includegraphics[width=0.6\columnwidth, trim=3cm 3cm 2.75cm 2.75cm, clip=true]{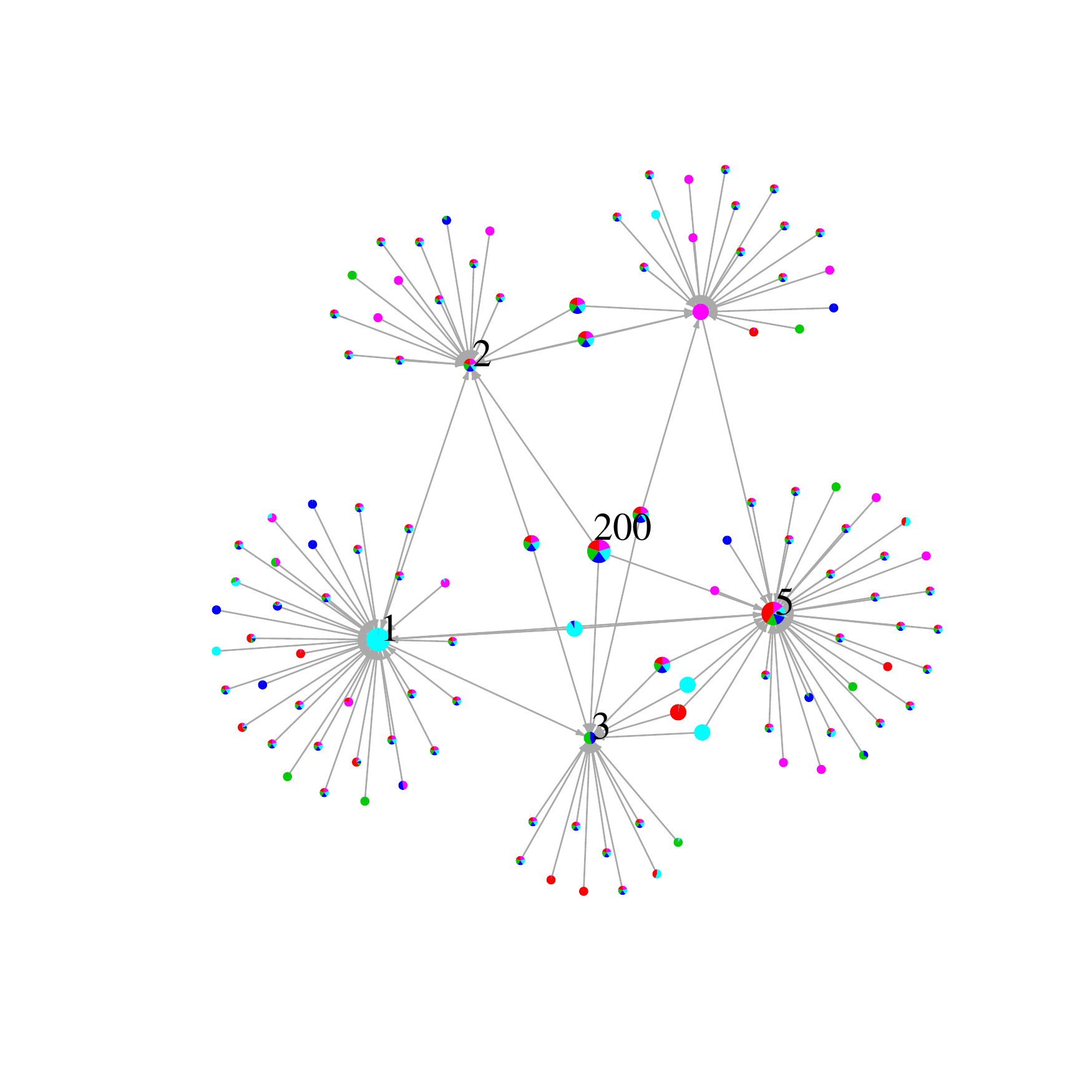}& 
\includegraphics[width=0.6\columnwidth, trim=3cm 3cm 2.75cm 2.75cm, clip=true]{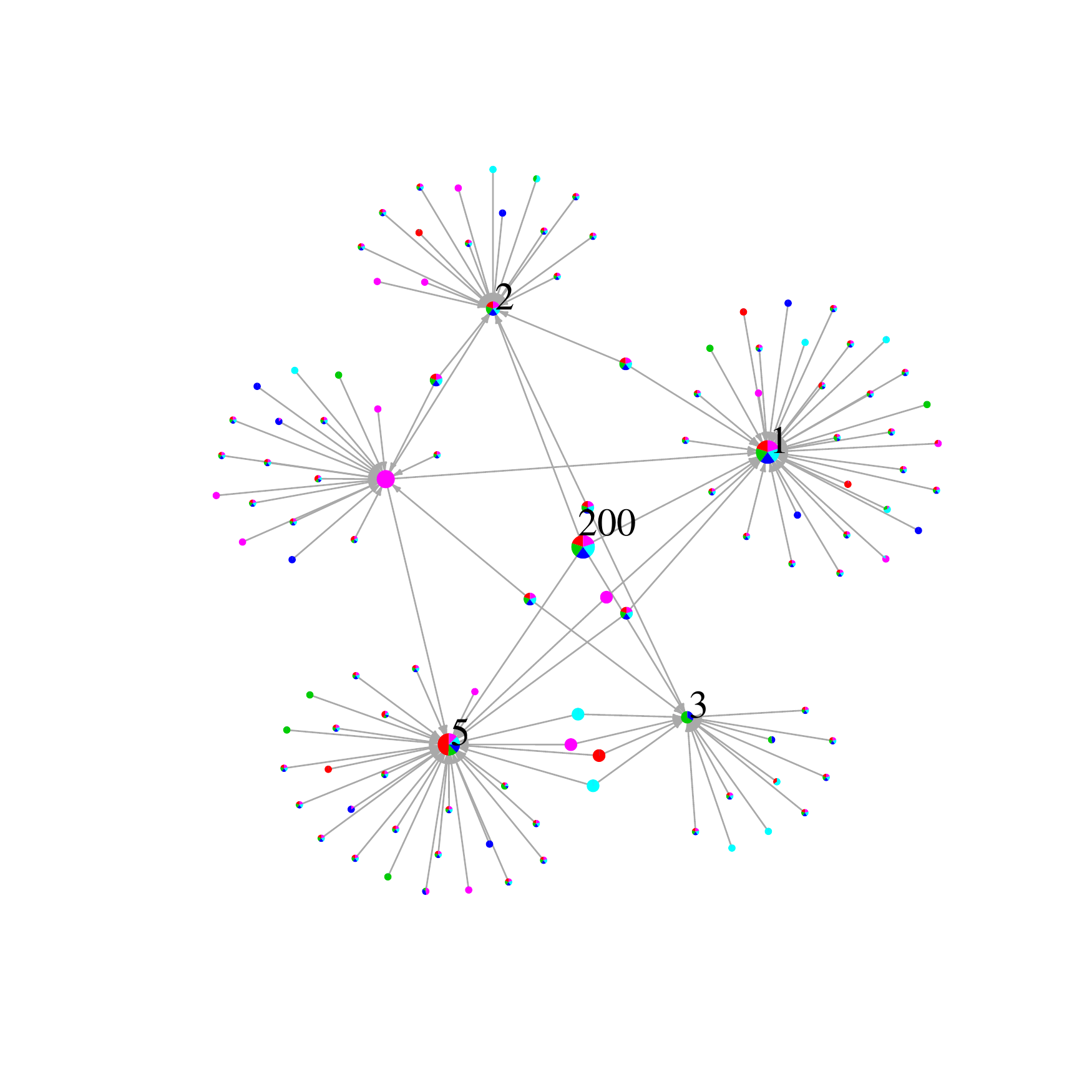}\\
\end{tabular}
}
\caption{(Color online) The raw (top row) and filtered Catalano networks (bottom row) colored by the $U_{t}$ community structure. A force directed layout in igraph was used to create this embedding. Nodes are colored by soft partitioning via the penalized NMF.}
\label{fig:vast-filtered}
\end{figure*}

\subsubsection{Catalano Communication Network}
The first example utilizes the Catalano social network, which was part
of the Visual Analytics Science and Technology (VAST) 2008 challenge
\cite{vast}. The synthetic data consists of 400 unique cell phone IDs
over a ten day period. Altogether, there are 9834 phone records with
the following fields: calling phone identifier, receiving phone
identifier, date, time of day, call duration, and cell tower closest
to the call origin. The purpose of the challenge was to characterize
the social structure over time for a fictitious, controversial
socio-political movement. In particular, the challenge requires
identifying five key individuals that organize activities and
communications for the network; a hint was given to challenge
participants that node 200 is one of the persons of interest. 

We use
the first seven days of data to illustrate our methodology, since
there is a strong change in the connection patterns from day 8-10 for
node 200 (see \cite{vast,vast-entry3} and references therein).
Directed networks are constructed daily by drawing an edge from the
caller to the receiver. Fig.~\ref{fig:vast-raw} shows an example of
one day's network. The graph is too cluttered to visually identify
leaders of the network or get a sense of the network structure.

We fit a sequence of rank 5 NMFs, as identified in Fig.~\ref{fig:vast}
through cross validation, with a large temporal penalty to highlight
the most persistent interactions. Fig.~\ref{fig:vast-filtered} shows
two sets of graph drawings for three days (due to space limitations),
with the nodes colored according to their community membership. The
first row shows the graph constructed directly from the data, while
the second row shows graph drawings of the \emph{fitted} model
$\hat{A}=U_{t}V_{t}^{T}$. The clustering results applied to the raw data 
are not interpretable, as the data is simply too cluttered. However, 
the persons of interest and the
hierarchical structure of the communication network are clearly shown
when considering the fitted networks. One can visually identify that
node $200$ consistently relays information to his neighbors (1,2,3,5),
who disseminate information to their respective subordinates.  We can
also see that nodes higher up on the organizational hierarchy tend to
belong to multiple communities, presumably since they disseminate
information to different groups of subordinates.

\begin{figure*}
\centerline{
\begin{tabular}{lccc}
 & Day 5 & Day 6 & Day 7 \\
\raisebox{3cm}{Raw} & \includegraphics[width=0.6\columnwidth, trim=3cm 3cm 2.75cm 2.75cm, clip=true]{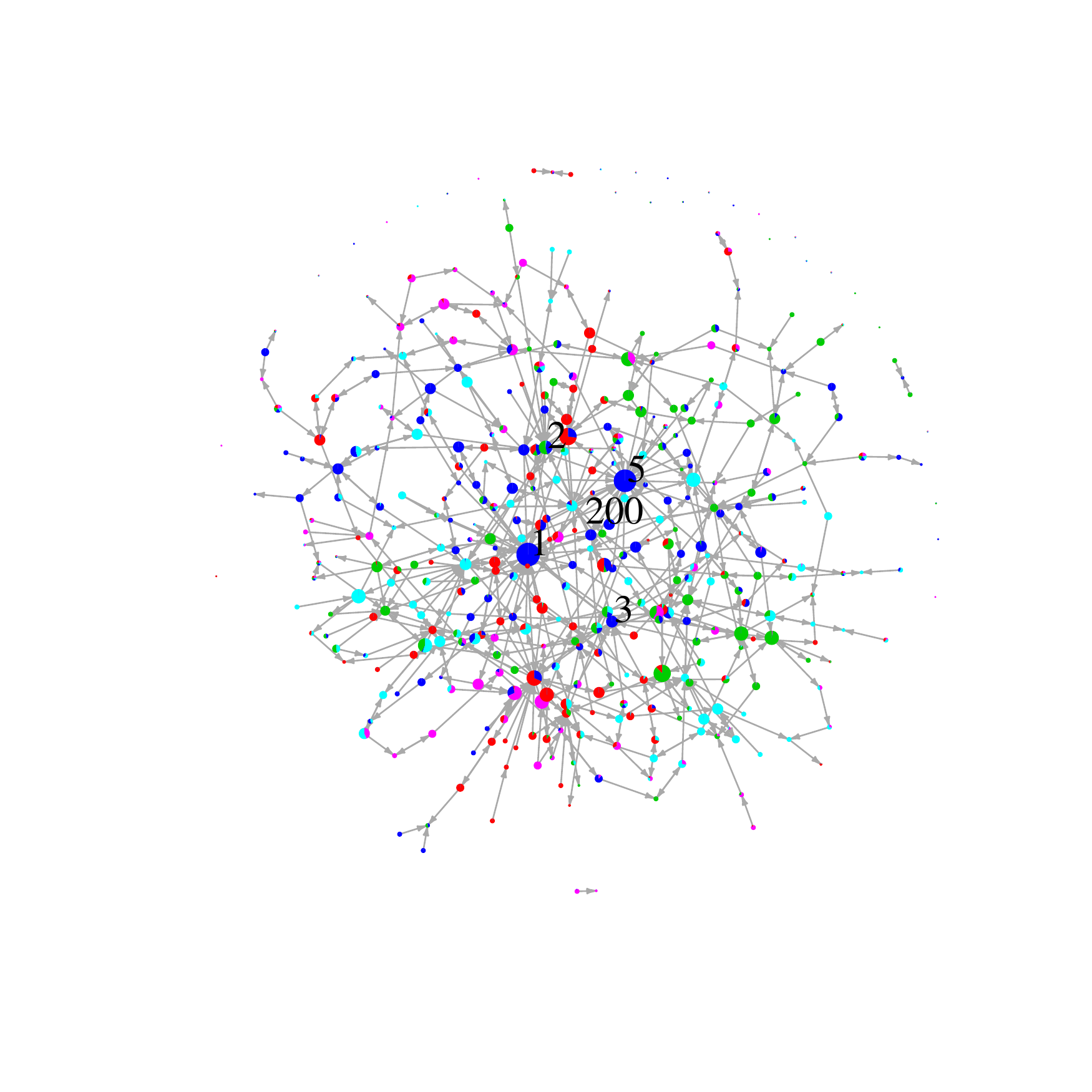}& 
\includegraphics[width=0.6\columnwidth, trim=3cm 3cm 2.75cm 2.75cm, clip=true]{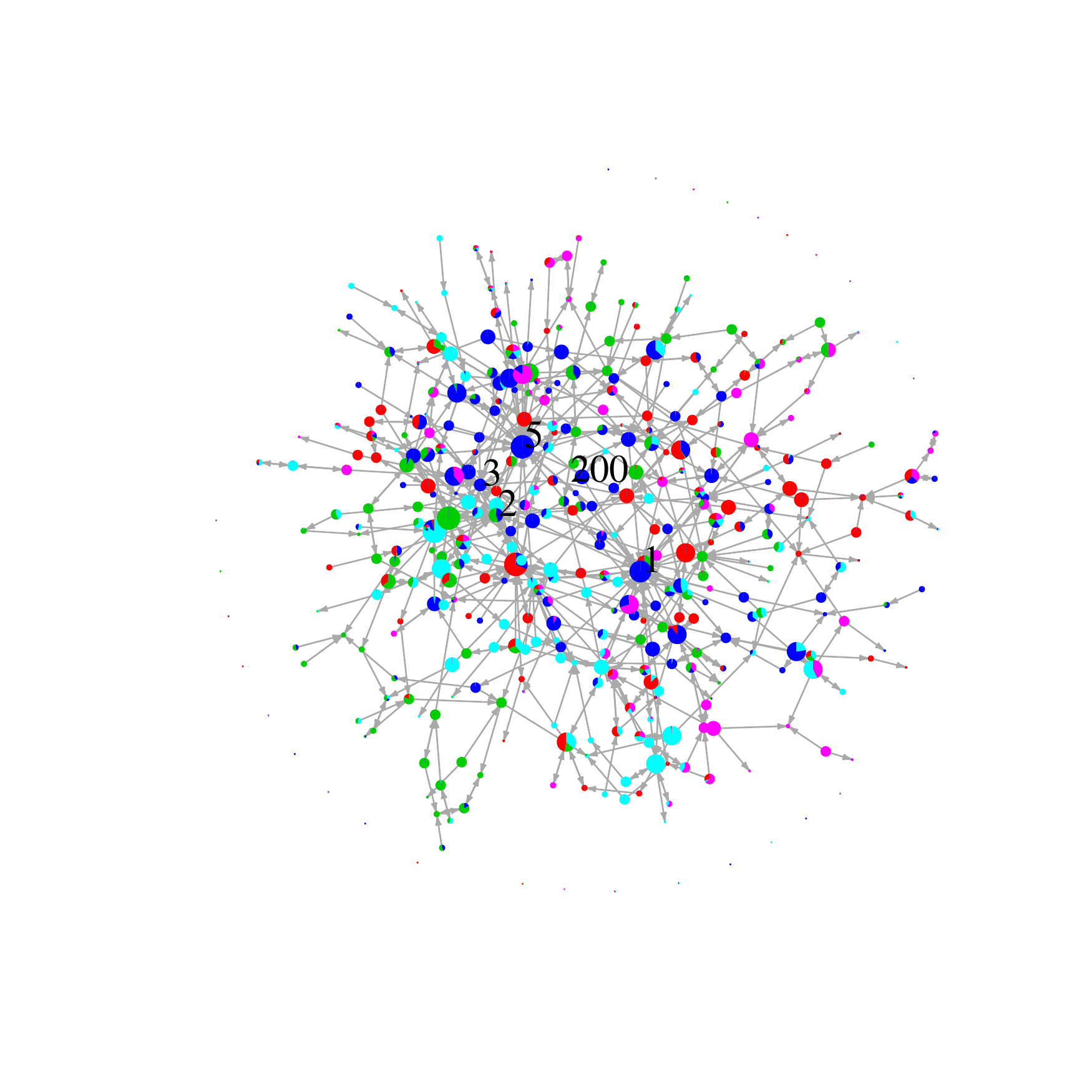}& 
\includegraphics[width=0.6\columnwidth, trim=3cm 3cm 2.75cm 2.75cm, clip=true]{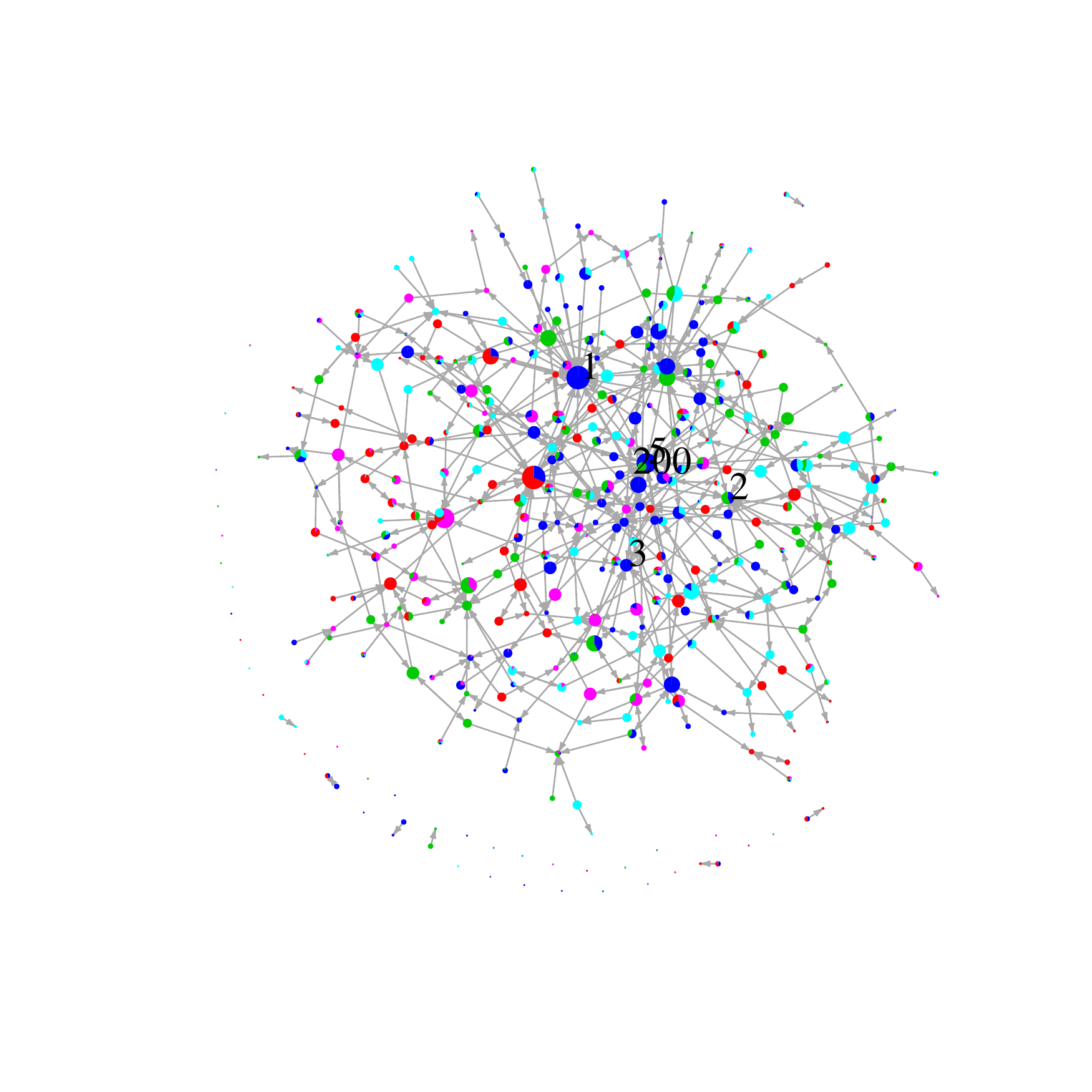}\\
\raisebox{3cm}{Fitted} & \includegraphics[width=0.6\columnwidth, trim=3cm 3cm 1.75cm 2.75cm, clip=true]{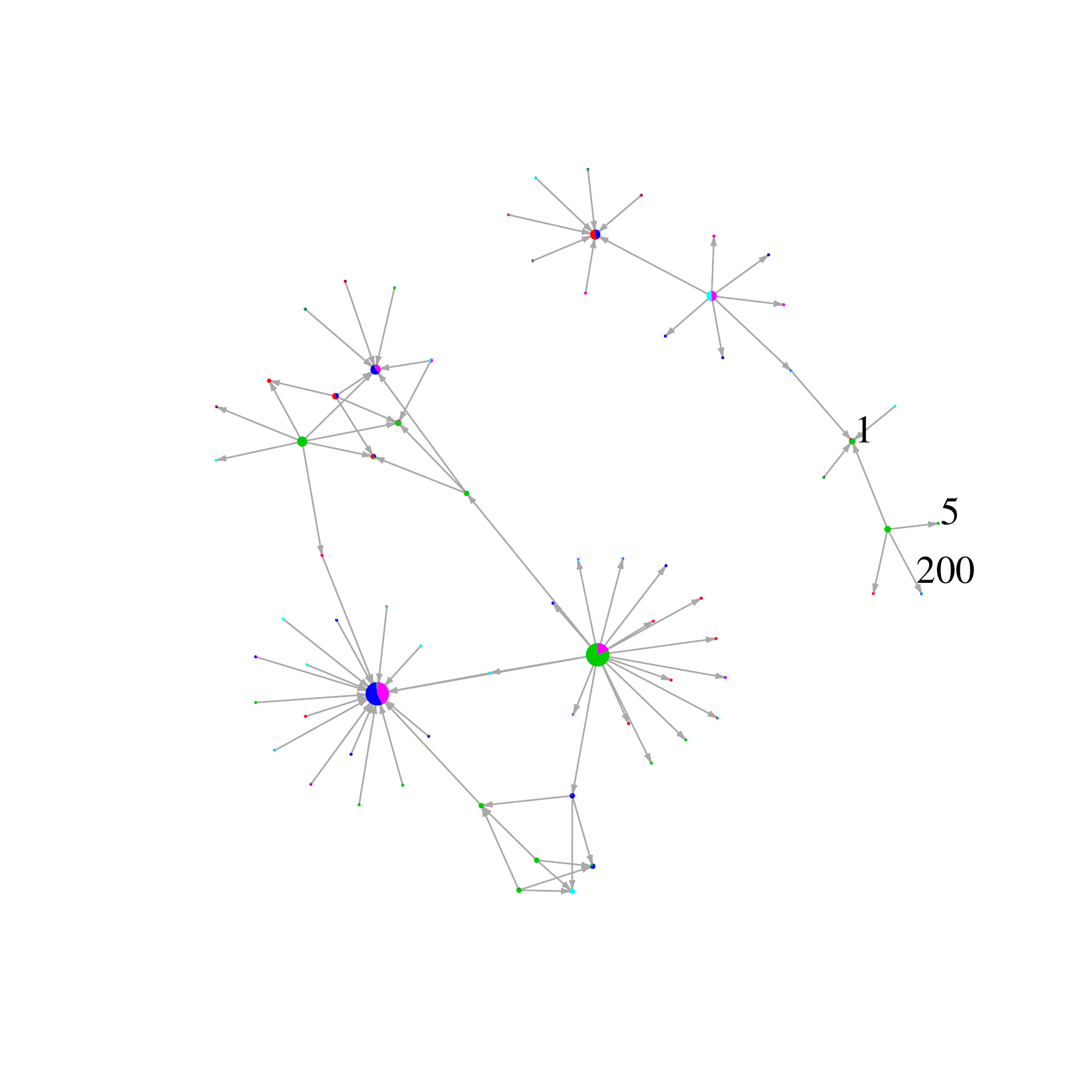}& 
\includegraphics[width=0.6\columnwidth, trim=3cm 3cm 2.75cm 2.75cm, clip=true]{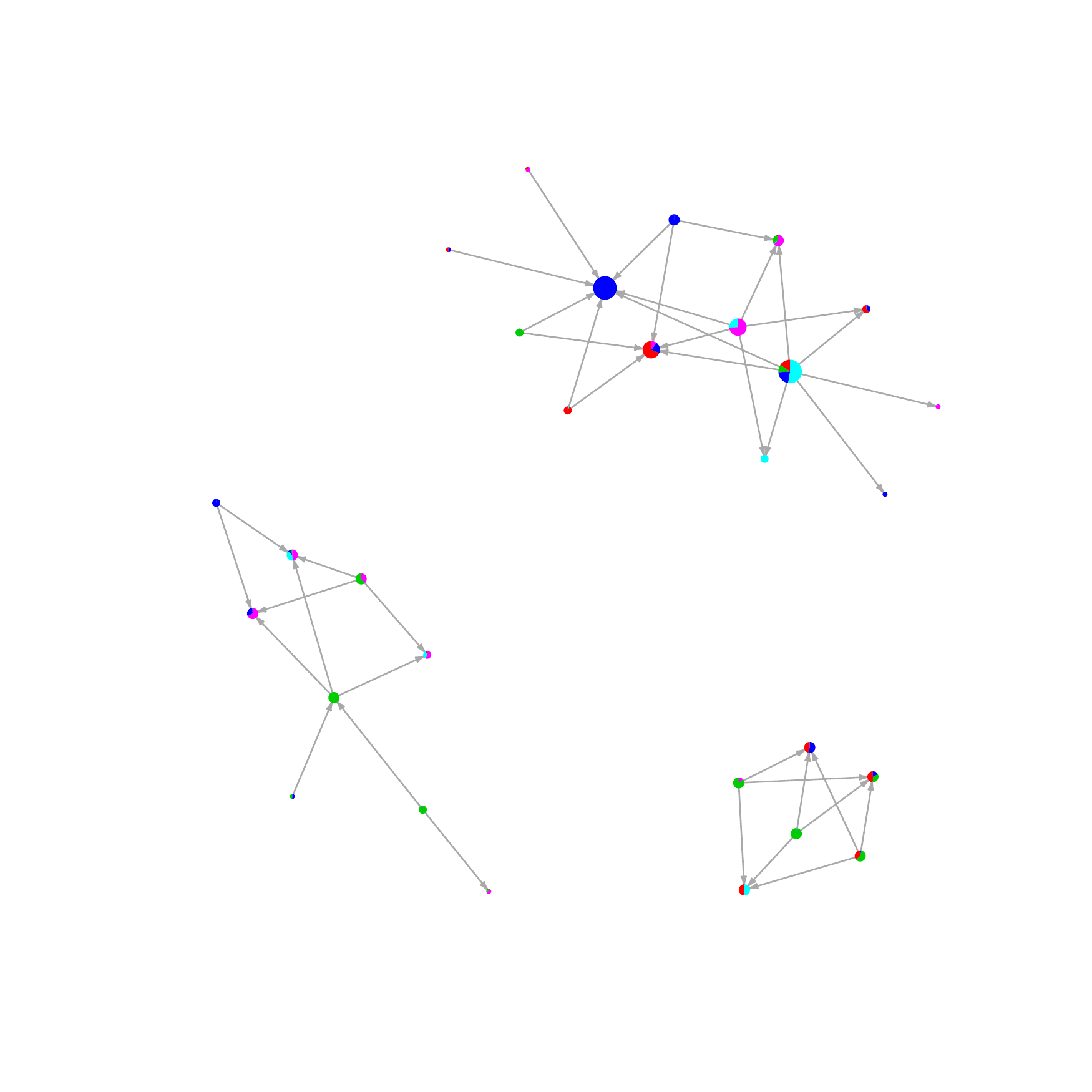}& 
\includegraphics[width=0.6\columnwidth, trim=3cm 3cm 2.75cm 2.75cm, clip=true]{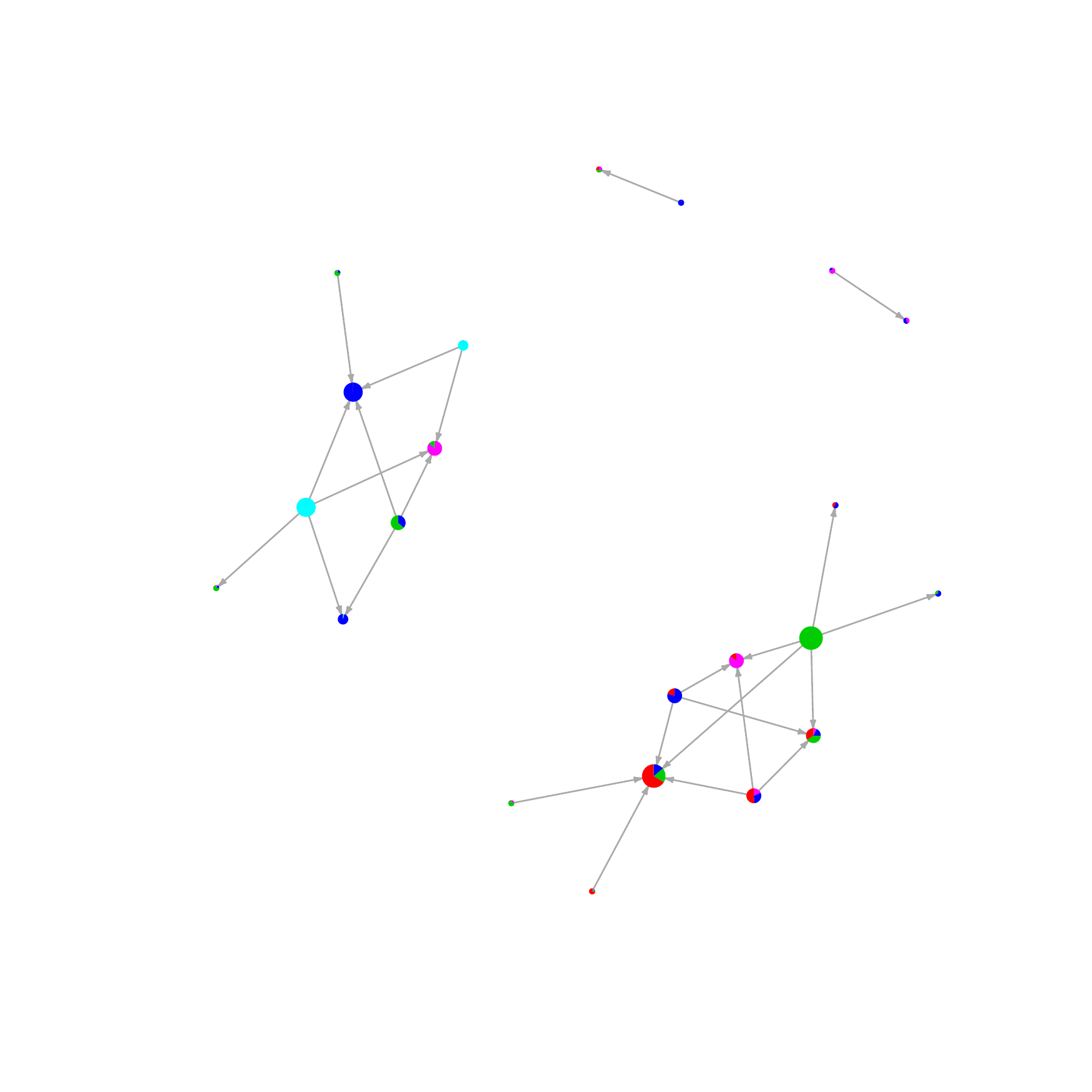}
\end{tabular}
}
\caption{(Color online) Results of applying the Facetnet factorization \cite{facetnet} with a prior weight of $\lambda=0.8$. The raw (top row) and filtered Catalano networks (bottom row) colored by the Facetnet factorization.}
\label{fig:vast-facetnet}
\end{figure*}

\begin{figure*}
\centerline{
\begin{tabular}{lccc}
 & Threshold=2 & Threshold=3 & Threshold=4 \\
\raisebox{3cm}{\begin{tabular}{l}Spectral\\Clustering\end{tabular}} & \includegraphics[width=0.6\columnwidth, trim=3cm 3cm 2.75cm 2.75cm, clip=true]{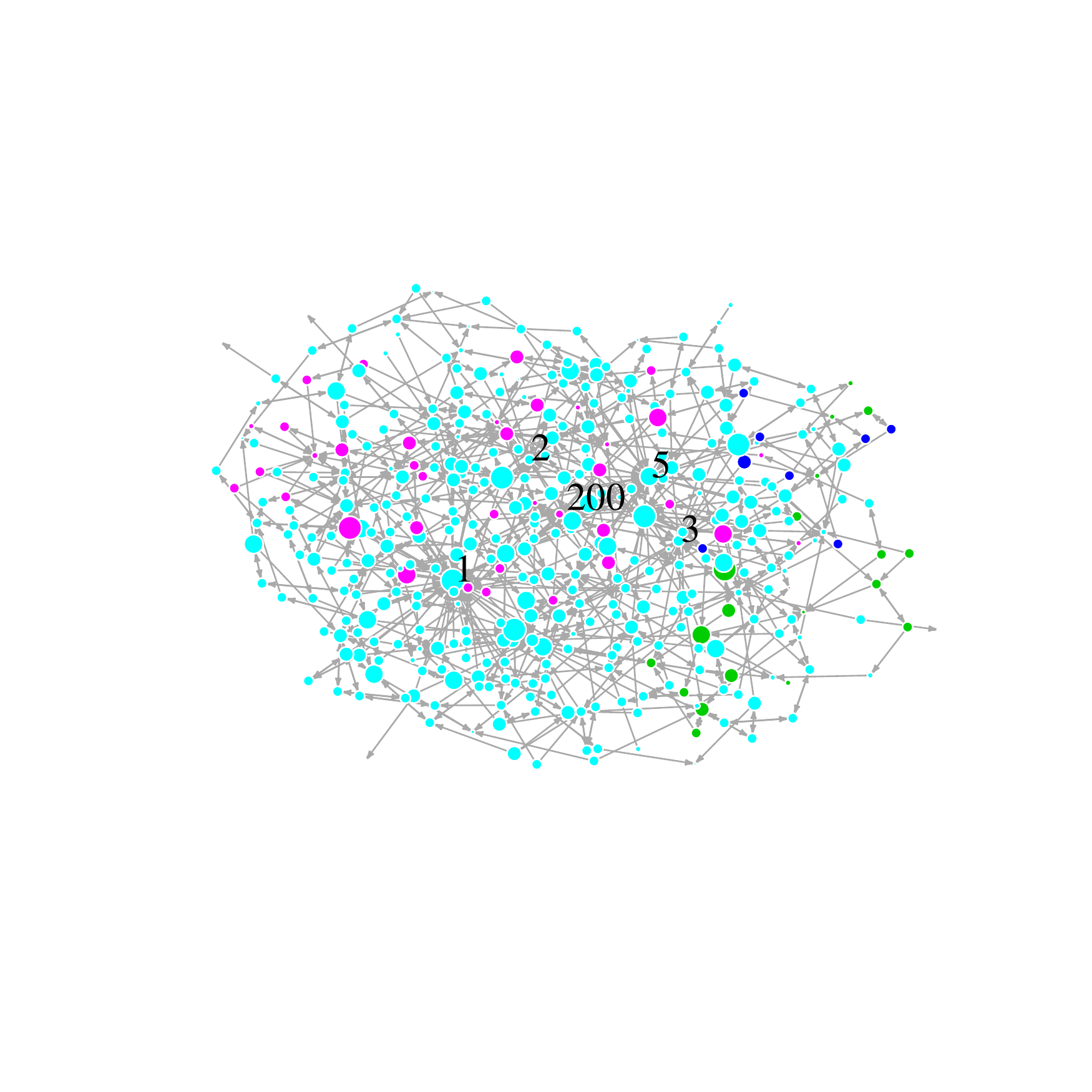}& 
\includegraphics[width=0.6\columnwidth, trim=3cm 3cm 2.75cm 2.75cm, clip=true]{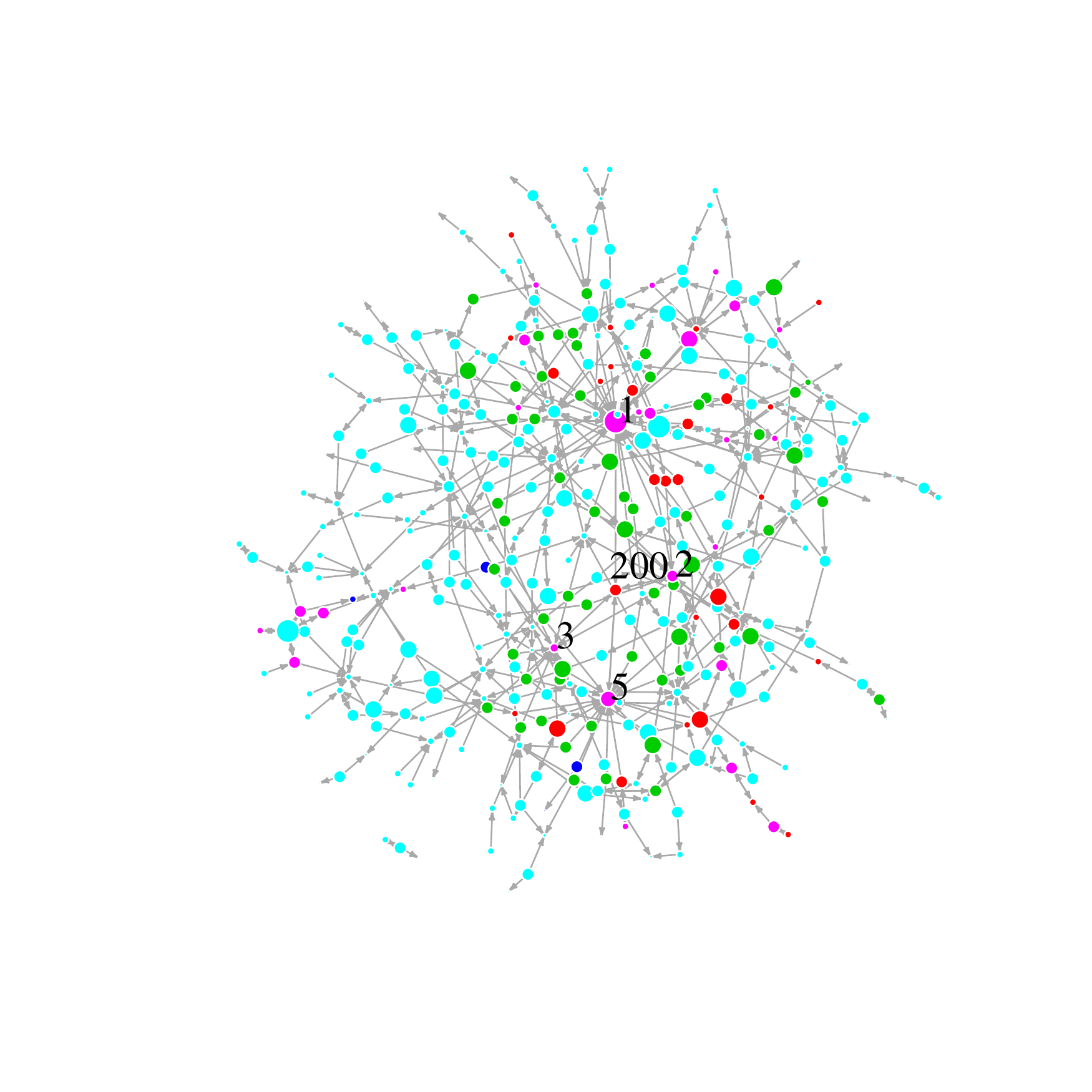}& 
\includegraphics[width=0.6\columnwidth, trim=3cm 3cm 2.75cm 2.75cm, clip=true]{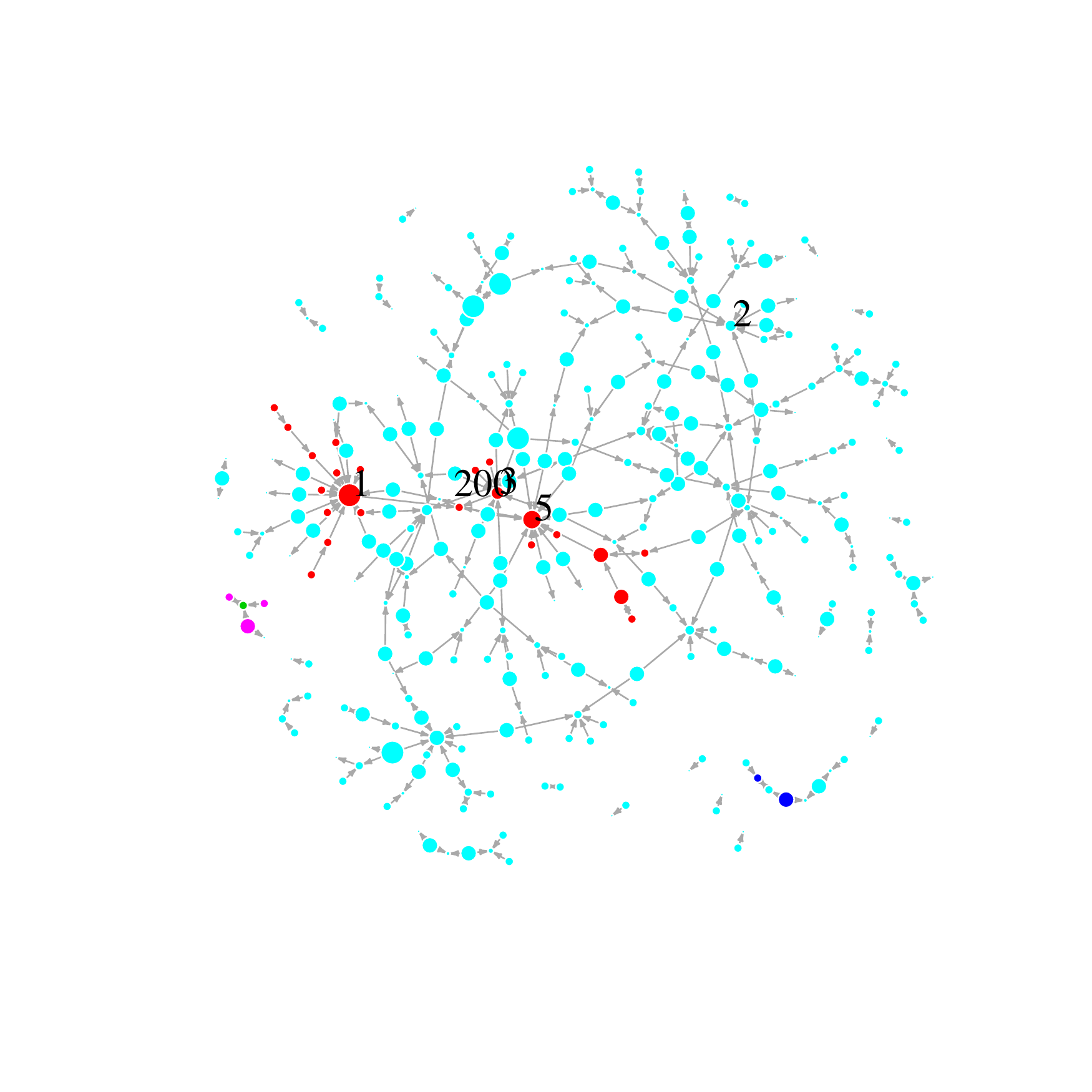}\\
\raisebox{3cm}{\begin{tabular}{l}Clique\\Percolation\end{tabular}} & \includegraphics[width=0.6\columnwidth, trim=3cm 3cm 2.75cm 2.75cm, clip=true]{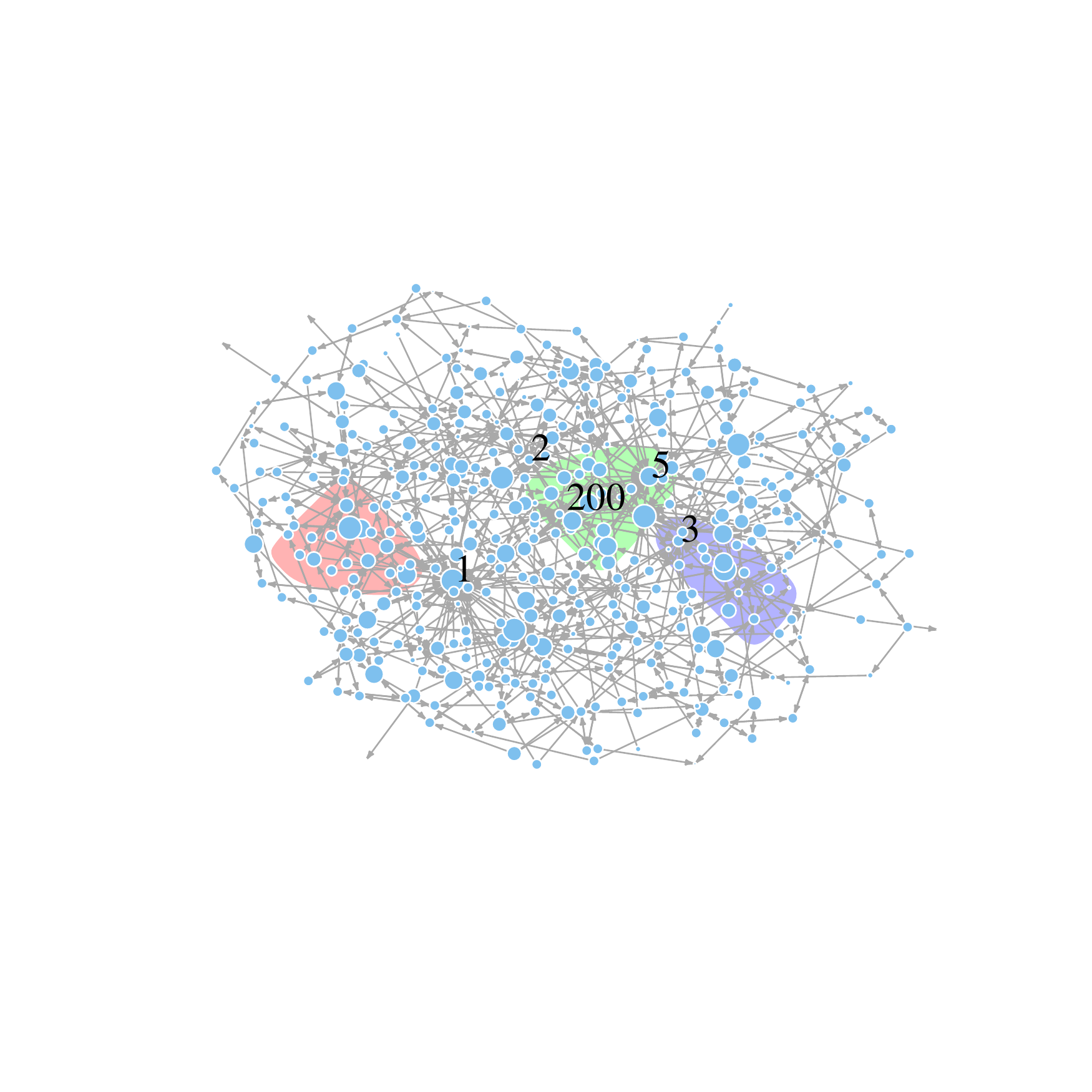}& 
\includegraphics[width=0.6\columnwidth, trim=3cm 3cm 2.75cm 2.75cm, clip=true]{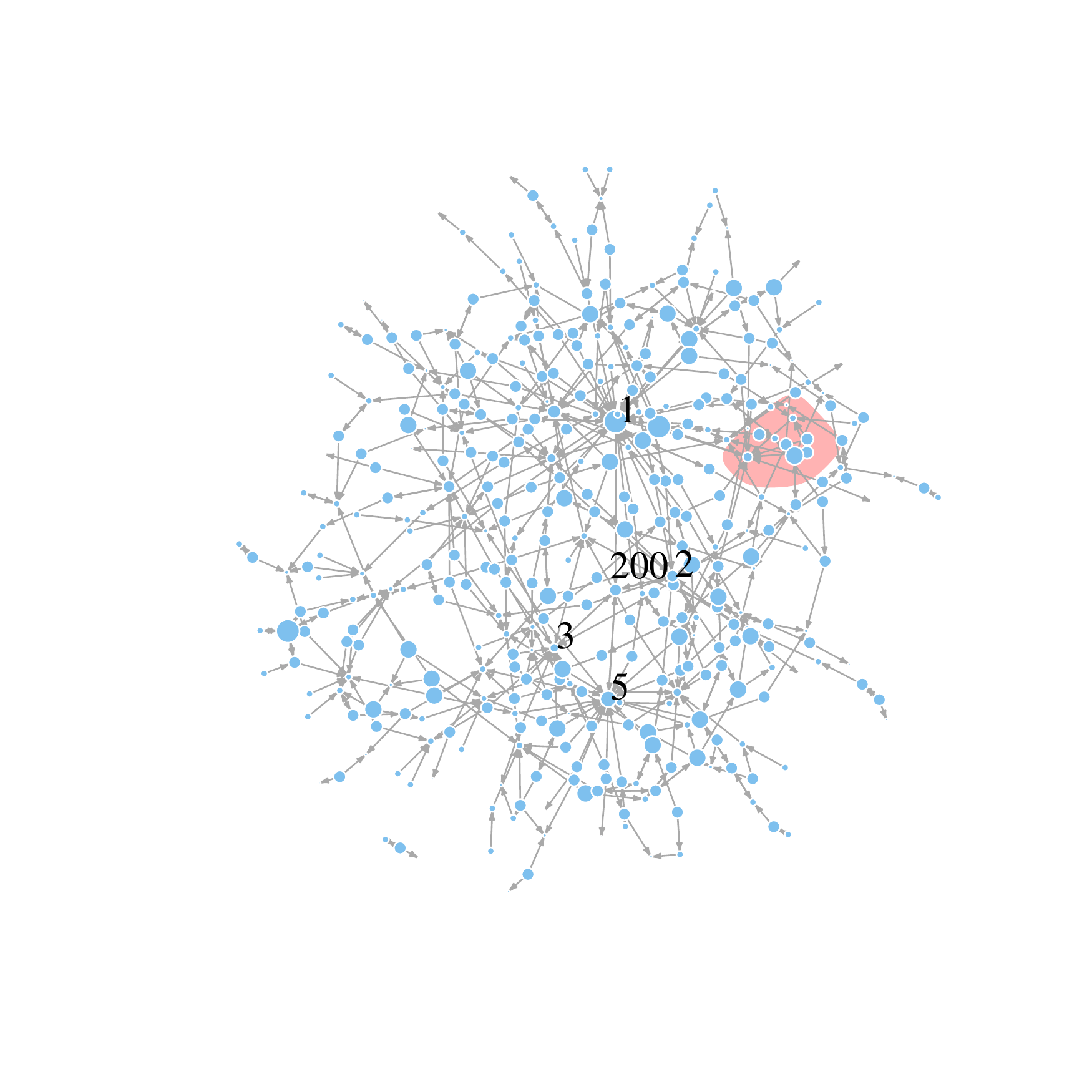}& 
\raisebox{3cm}{No Communities Detected}
\end{tabular}
}
\caption{(Color online) The first and second rows apply static clustering methods to the collapsed data (averaging over time). All alternative methods struggle to identify the key individuals or hierarchical organizational structure.}
\label{fig:vast-static}
\end{figure*}

Fig.~\ref{fig:vast-facetnet} shows the results of applying 
Facetnet \cite{facetnet}, an alternative NMF methodology for
dynamic overlapping community detection. Facetnet applies an 
underlying model with less flexibility resulting in poor reconstructions of the 
data, as seen in the fitted networks. We also collapse the data
into a single network snapshot in order to apply static clustering
algorithms. First, an edge is kept only if it was observed more than
\emph{Threshold} days. Then, spectral clustering and clique
percolation are applied to the resultant network snapshot.  All
alternative methods struggle, as the data is too `hairball' like. On
the other hand, the fitted penalized NMF model provides a unified
framework to filter the network and visualize community
structure.  VAST never officially released correct answers for the
challenge. However, our analysis closely matches winning entries
\cite{vast-entry, vast-entry2, vast-entry3}. Treating the conclusions
of the entries as ground truth, we have provided a simple workflow
that uncovers patterns in the data that are not directly obtainable
with traditional methods.

\subsubsection{Preferential Attachment Process}
\begin{figure} 
\centerline {
\subfigure[No penalties] {
\includegraphics[width=.5\columnwidth]{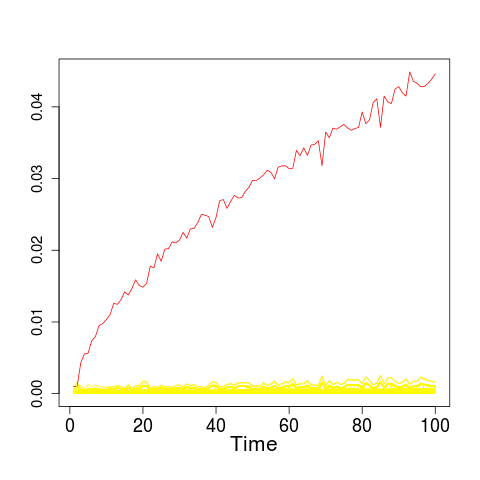}
\includegraphics[width=.5\columnwidth]{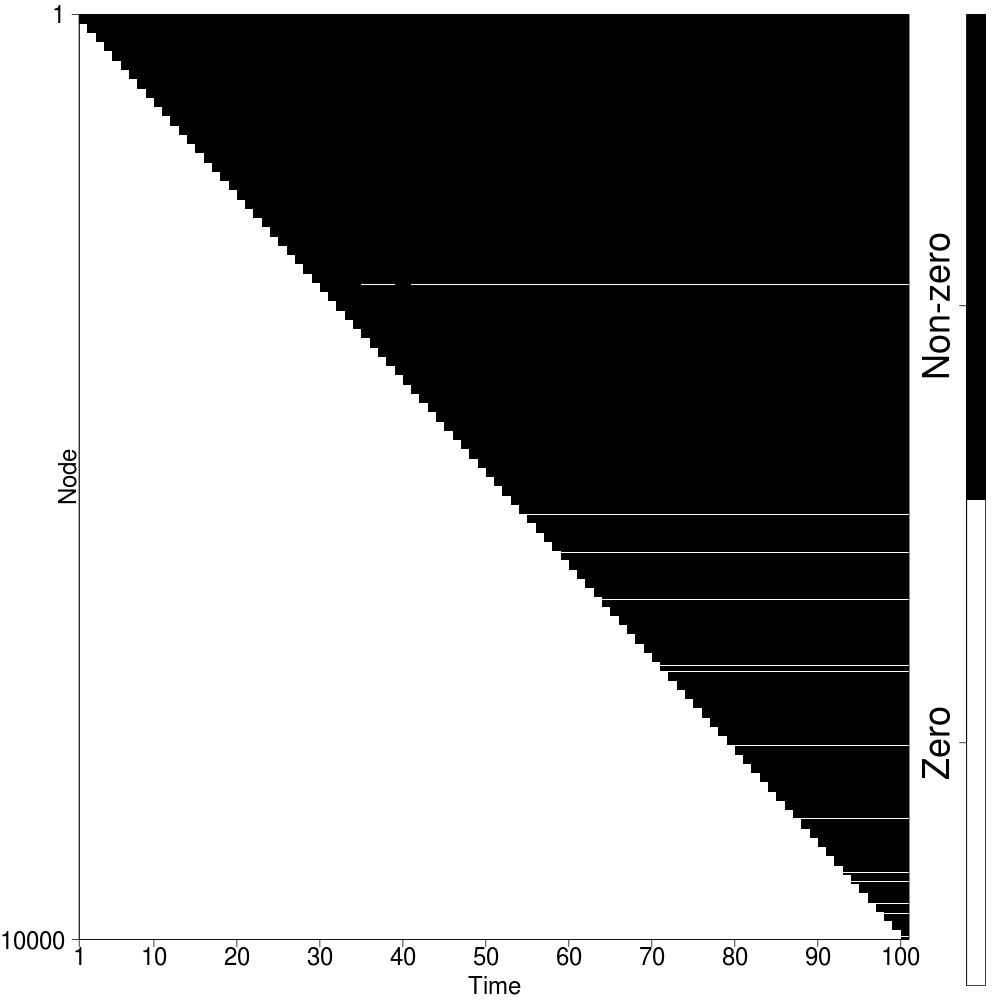}
}
}
\centerline {
  \subfigure[$\lambda_t = 50, \lambda_s = 5$] {
\includegraphics[width=.5\columnwidth]{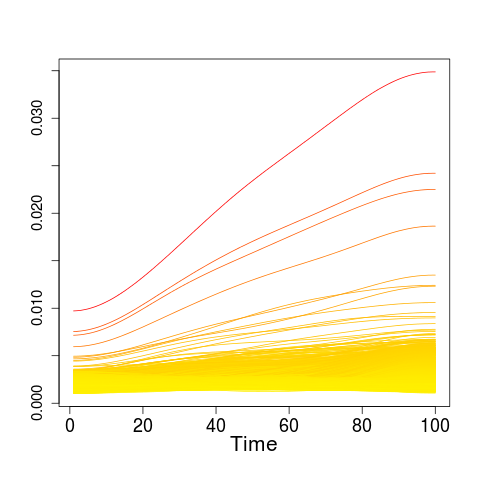}
\includegraphics[width=.5\columnwidth]{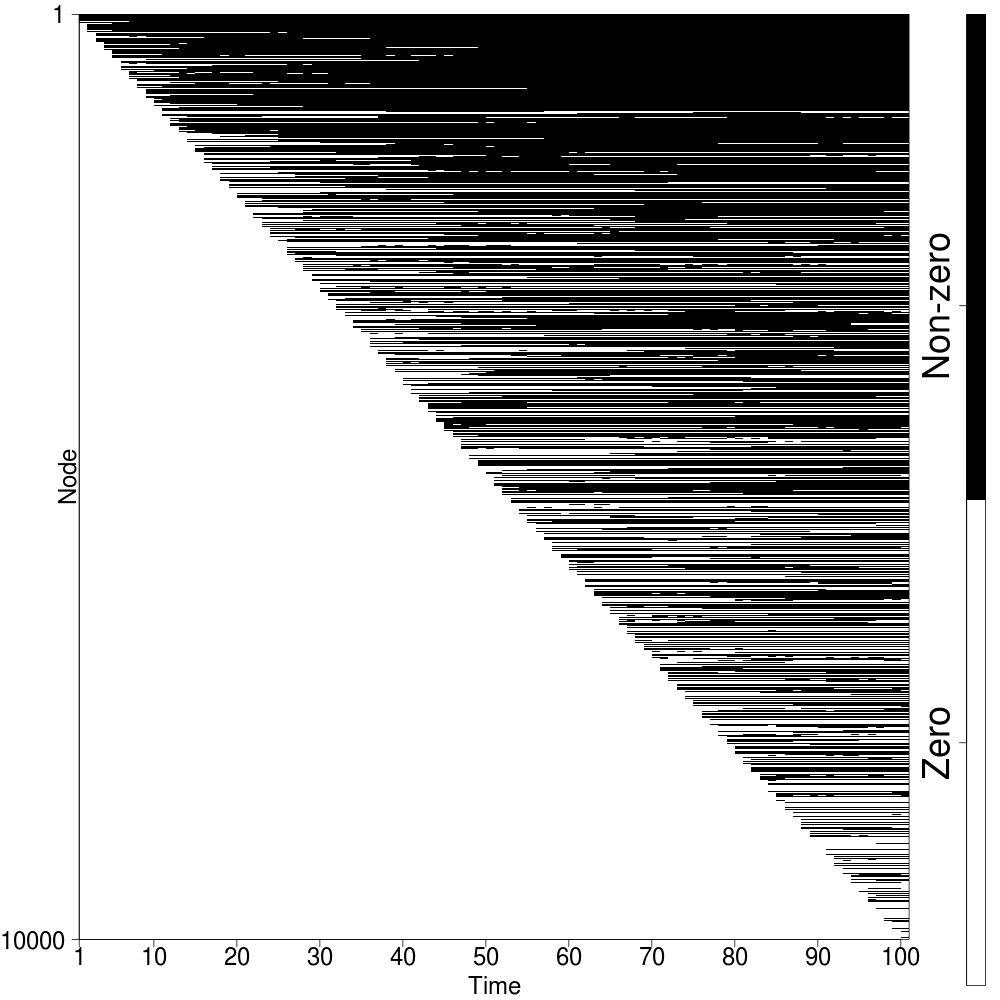}
}
}
\centerline {
  \subfigure[$\lambda_t = 100, \lambda_s = 5$] {
\includegraphics[width=.5\columnwidth]{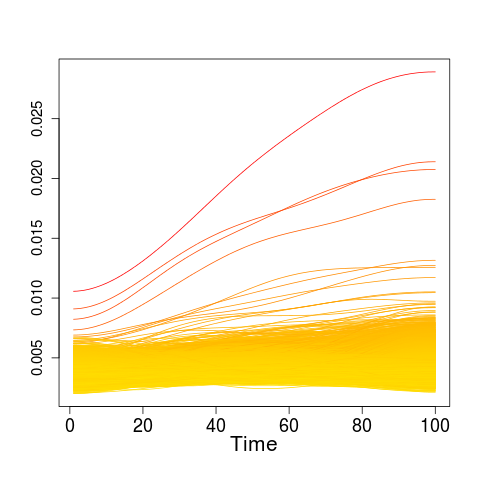}
\includegraphics[width=.5\columnwidth]{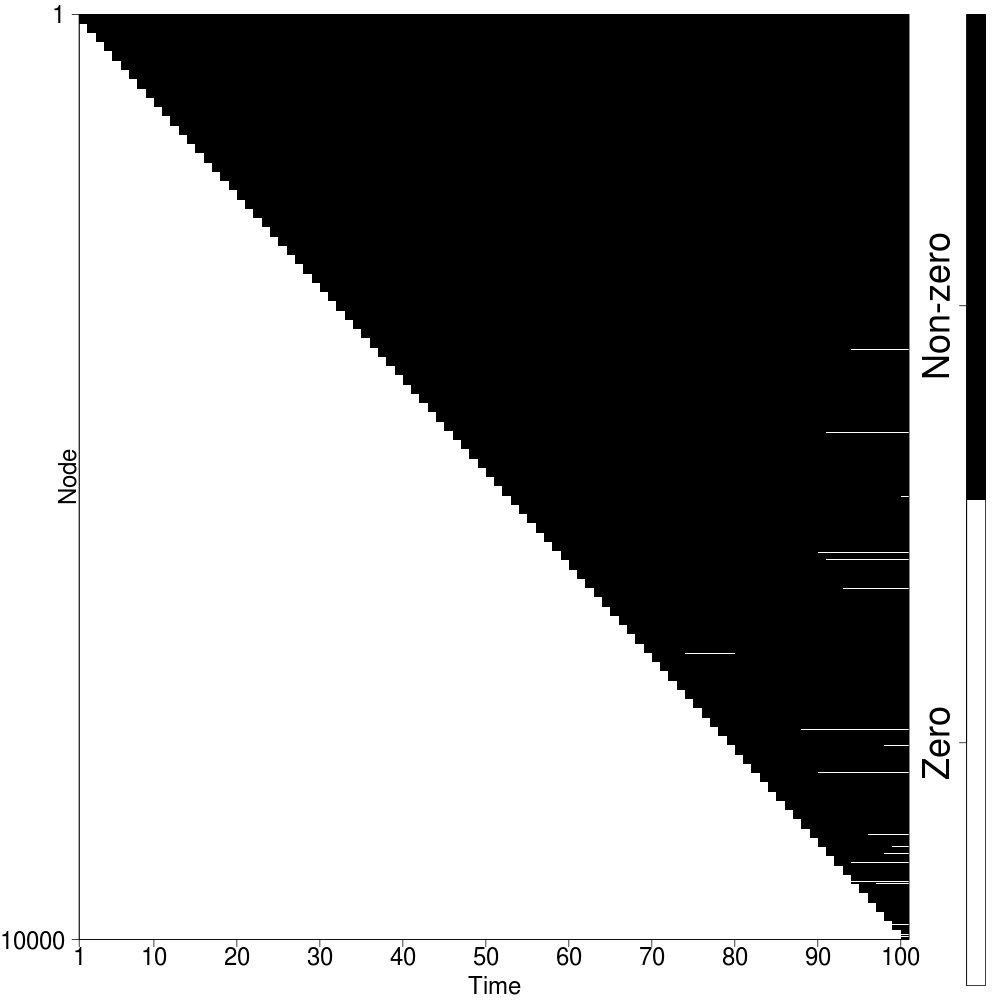}
}
}
\caption{(Color online) Fitted values for $U$ and $V$ over time for
  the preferential attachment simulation. The left column shows a time
  plot of $U_t$ over different parameter values. Each line corresponds
  to a node on the graph. The right column identifies the nonzero
  elements of $V_t$. Each row corresponds to a node on the graph and
  time varies along the horizontal axis.}
\label{fig:clustExpressionsPA}
\end{figure}

In this simulation, nodes attach according to a preferential
attachment model \cite{newman2006structure,Barabasi} until $10000$
nodes have 'attached' to the embedding. We observe this growing
process at 100 uniformly spaced time points. Thus, at each time point
100 new nodes attach to the graph. We use source code from a networks
MATLAB toolbox \cite{toolbox} that generates preferential attachment
graphs according to the standard model.

In the preferential attachment model, $\Pi(i)$, which represents the
probability that a new node connects to node $i$, depends on node
$i$'s degree. Specifically, we have
\begin{equation}
\Pi(i) \propto d_{i} 
\end{equation}
where $d_{i}$ is the degree of the $ith$ node. 
This generating framework leads to networks
whose asymptotic degree distribution follows a power-law distribution with
parameter $\gamma=3$. Graphs with heavy-tailed degree distributions
are commonly observed in a variety of areas, such as the Internet,
protein interactions, citation networks, among others \cite{powerLaw}.

In practice, an analyst would not know that the data comes 
from a preferential attachment process. In which case, an 
exploratory analysis may include inspecting the network sequence on a set of standard metrics 
(degree, transitivity, centrality, etc.), graph drawings, as well as community detection
approaches. 
We believe that a sequence of one-dimensional ($K=1$) penalized NMFs can serve as the basis
for a complimentary exploratory tool that helps uncover different connectivity patterns and 
evolution in the data. In particular, due to the smoothness penalty, 
time plots in $U_{t}$ for each node become useful for uncovering the number and types of 
node evolutions in the data. Similarly, heatmaps or displays of the sparsity 
pattern of $V_{t}$ are useful to identify when nodes/groups become significantly active. 

Since preferential attachment networks have been extensively studied, we show only the NMF-based 
displays. Fig.~\ref{fig:clustExpressionsPA} shows important (hub) nodes that distinct trajectories
 that indicate their increasing importance to the network over time. 
The $V_t$ sparsity features a pseudo-upper triangular form. This corresponds to the node
attachment order and reflects that nodes permanently attach after 
connecting to the network. Such displays can be created quickly and can help
 the process of identifying interesting nodes, formulating research questions, and so on.

Also shown in Fig.~\ref{fig:clustExpressionsPA} is that penalization is important to the 
usefulness and interpretability of the displays. For instance, without a
temporal penalty, the time plots emphasize only the highest degree node. 
With appropriate penalties, an analyst can visually identify the different hub nodes.

\subsection{Real Networks}

\subsubsection{arXiv Citations}
We investigate a time series of citation networks provided as part of
the 2003 KDD Cup \cite{KDD-2003}. The graphs are from the e-print
service arXiv for the `high energy physics theory' section.

The data covers papers in the period from October 1993 to December
2002, and is organized into monthly networks. In particular, if paper
$i$ cites paper $j$, then the graph contains a directed edge from $i$
to $j$. Any citations to or from papers outside the dataset are not
included. Following convention, edges are aggregated, that is, the
citation graph for a given month will contain all citations from the
beginning of the data up to, and including, the current
month. Altogether, there are $22750$ nodes (papers) with $176602$
edges (references) over $112$ months.

\begin{figure}
\centerline{
\begin{tabular}{ccc}
Jan 1995 & Jan 1998 & Jan 2000\\
\includegraphics[width=.33\columnwidth]{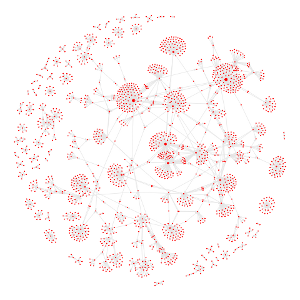} &
\includegraphics[width=.33\columnwidth]{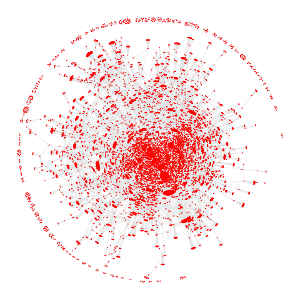} &
\includegraphics[width=.33\columnwidth]{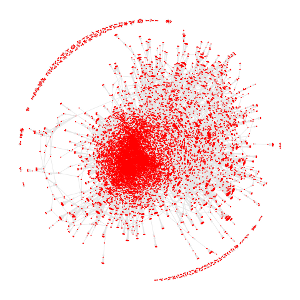} 
\end{tabular}
}
\caption{Graph layouts of the arXiv data at three different time points. Due to the size of the networks, it quickly becomes difficult to discern paper (node) properties.}
\label{fig:arXiv:rawlayouts}
\end{figure}

\begin{figure}
\centerline{
\includegraphics[width=.5\columnwidth]{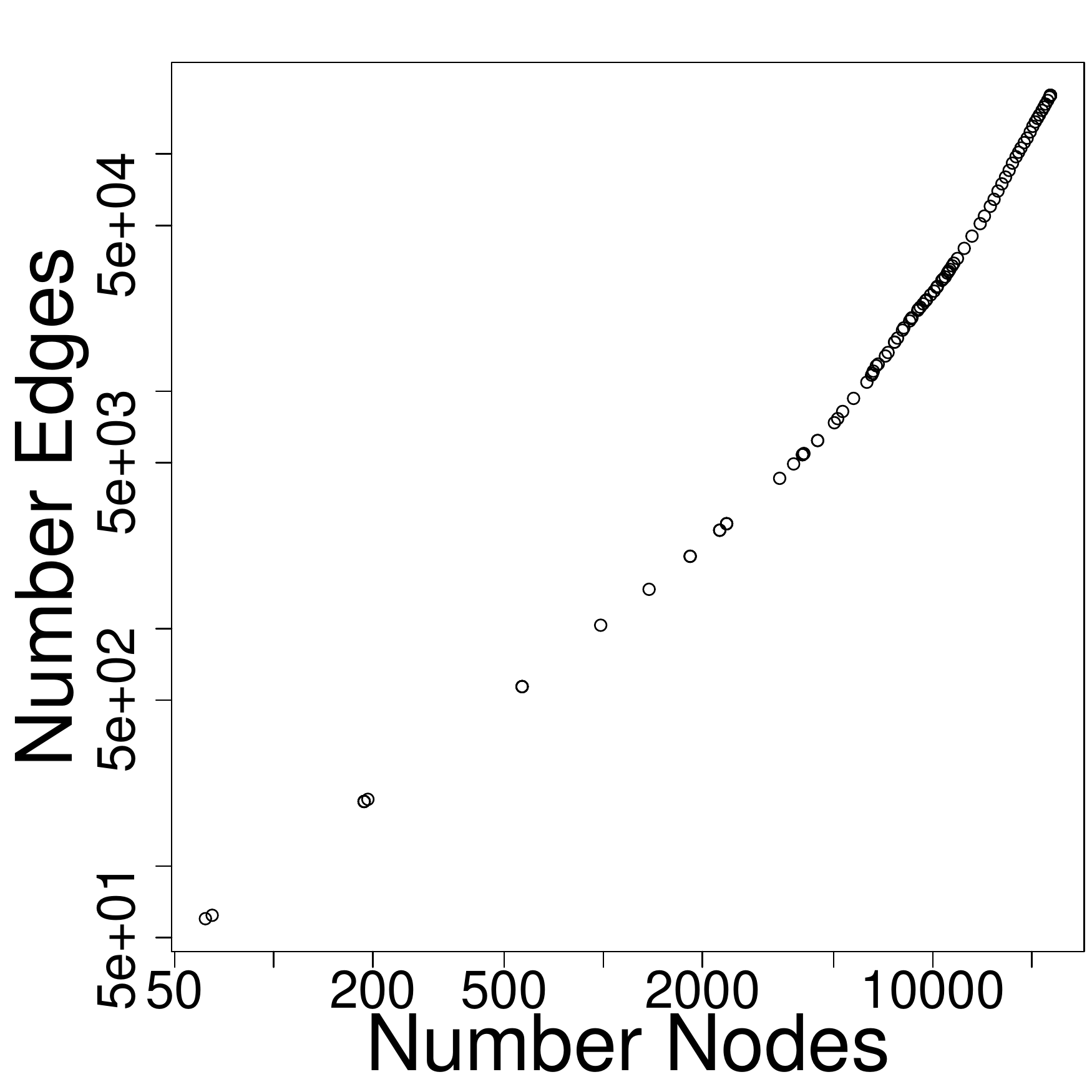} 
\includegraphics[width=.5\columnwidth]{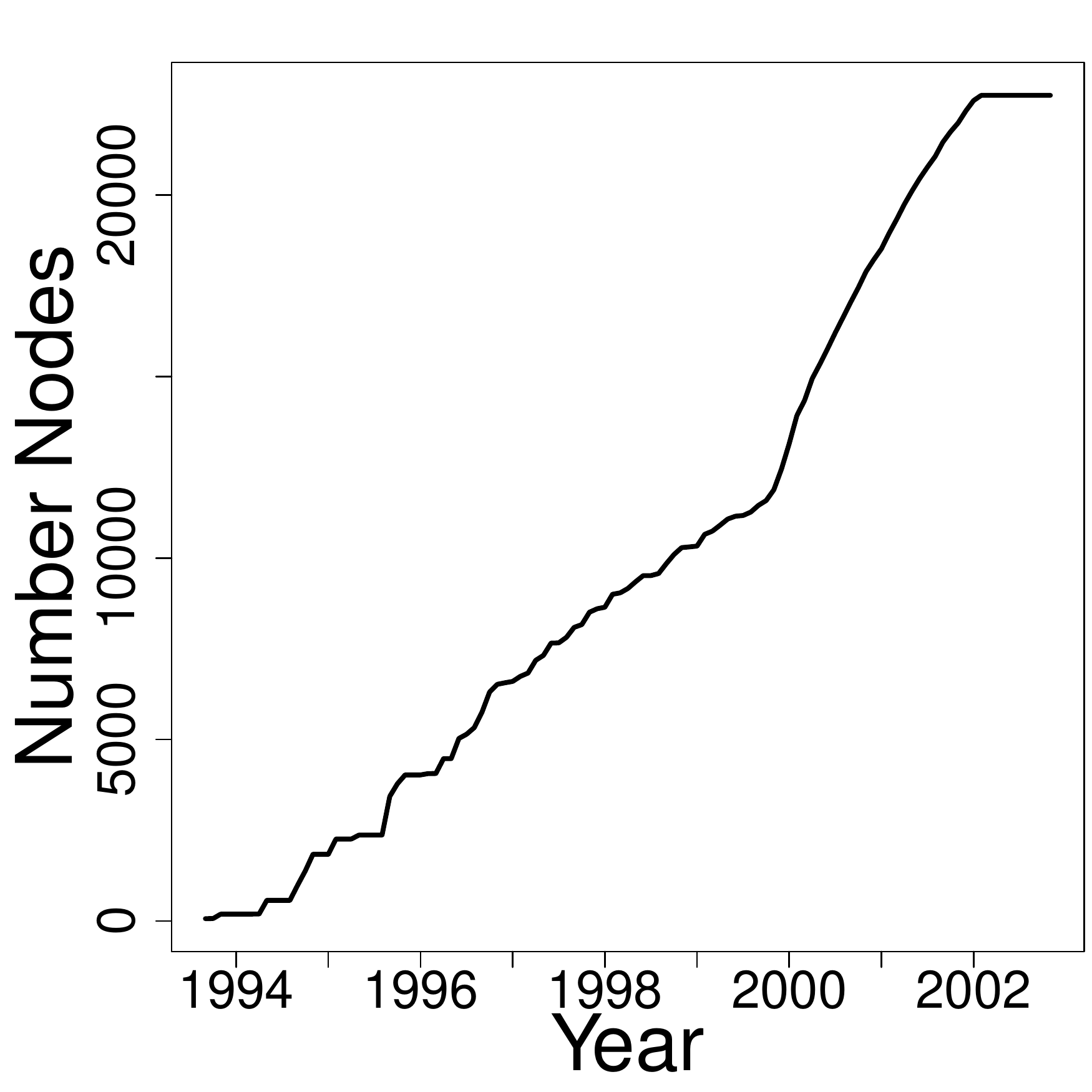}
}
\centerline{
\includegraphics[width=.5\columnwidth]{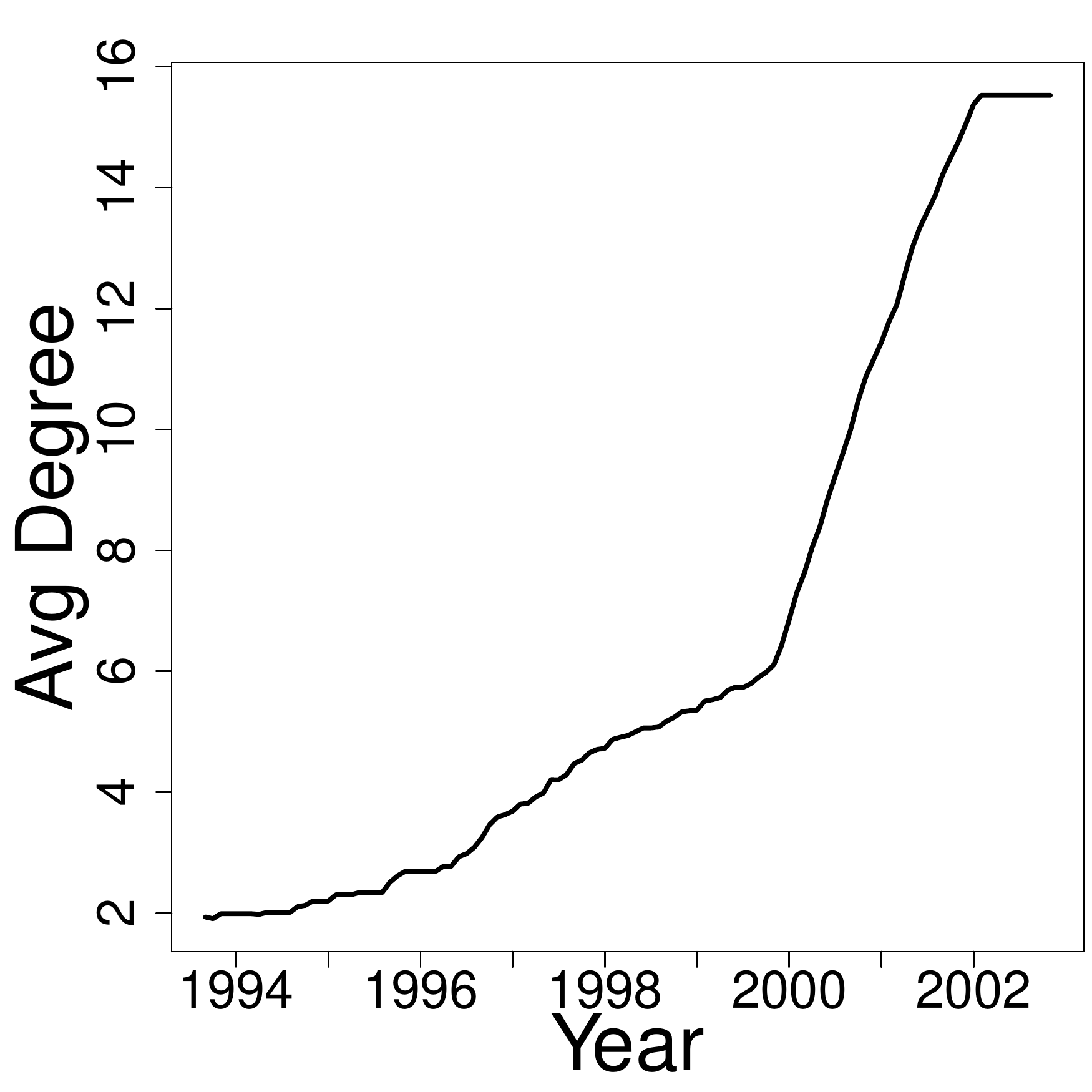}
\includegraphics[width=.5\columnwidth]{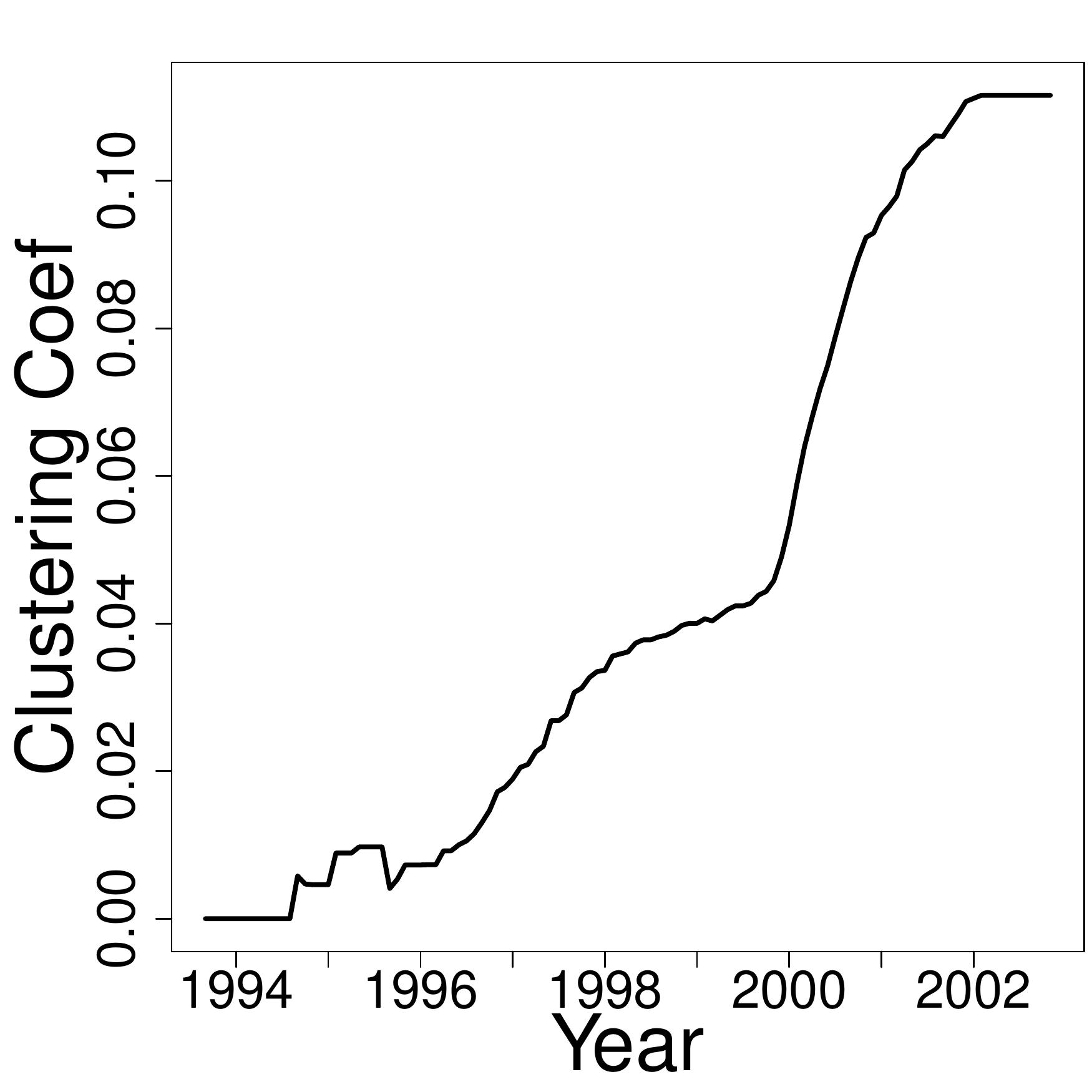}
}
\caption{(Color online) ArXiv network statistics over time. The kink near
  near January 2000 indicates sudden, rapid growth. The top-left plot
  is on a log-log scale.}
\label{fig:arXiv:basic}
\end{figure}

As a first step towards investigating the data, we draw the network at 
different points in time in Fig.~\ref{fig:arXiv:rawlayouts}.
Even when considering a single time point, it quickly becomes difficult 
to discern paper (node) properties due to the large network size. 
Thus, the data requires network statistics and other methods 
to extract structure and infer dynamics in the network sequence. 
Network 
statistics, shown in Fig.~\ref{fig:arXiv:basic}, provide some additional 
insight. There is a noticeable 
increase in network growth around the year 2000, which is 
commonly attributed to 
papers that reference other works before the start of the observation
period (see \cite{Leskovec-graphs-time}). As we move away from the
beginning of the data, papers primarily reference other papers
belonging to the data set. Additional statistical properties of the
data were discussed in \cite{Leskovec-graphs-time}, which found that
the networks feature decreasing diameter over time and heavy-tailed
degree distributions.

\begin{figure*}
\begin{tabular}{cccc}
No Penalty & $\lambda_s = 1$ & $\lambda_s = 5$ & $\lambda_s = 10$
\\ \includegraphics[width=.5\columnwidth]{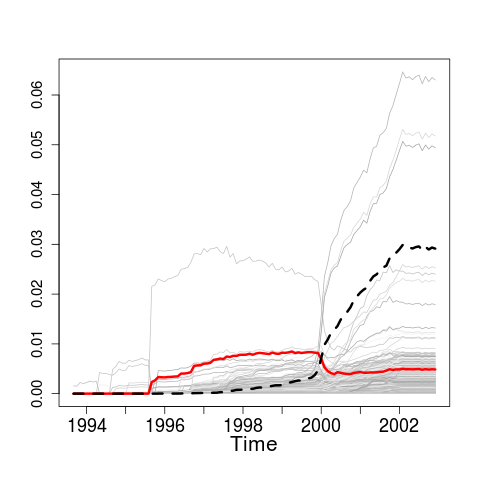}
&
\includegraphics[width=.5\columnwidth]{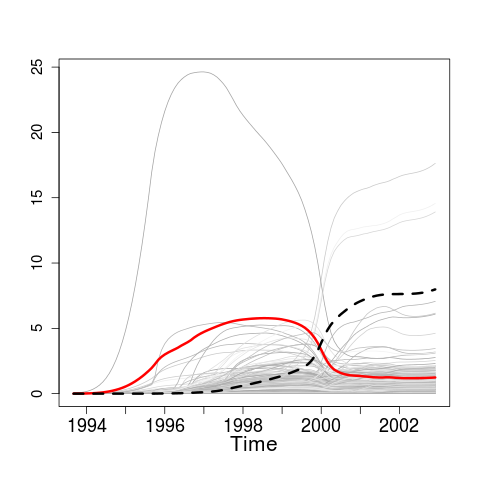}
&
\includegraphics[width=.5\columnwidth]{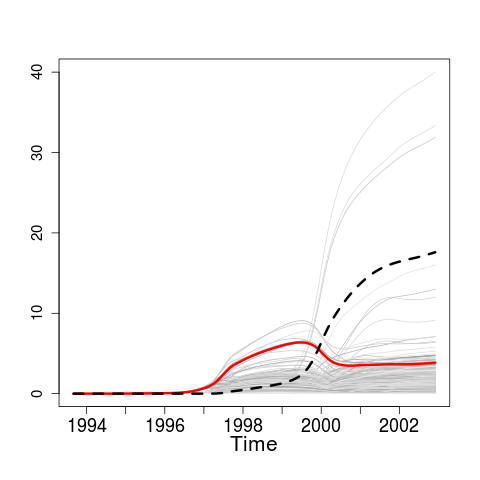}
&
\includegraphics[width=.5\columnwidth]{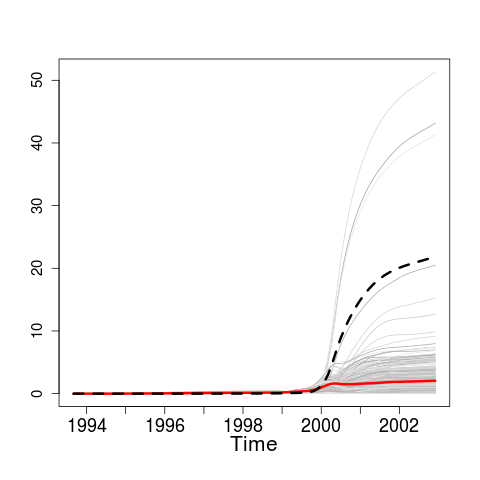}
\\ \includegraphics[width=.5\columnwidth]{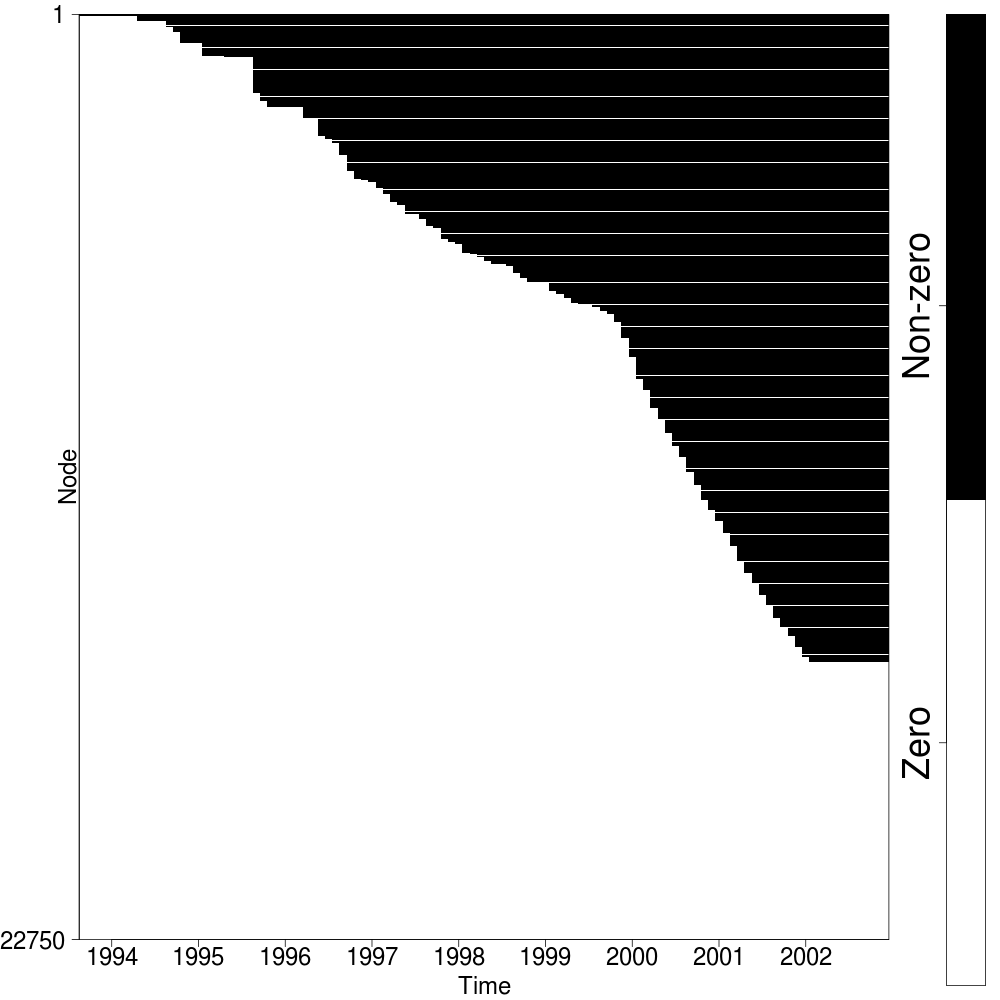}
&
\includegraphics[width=.5\columnwidth]{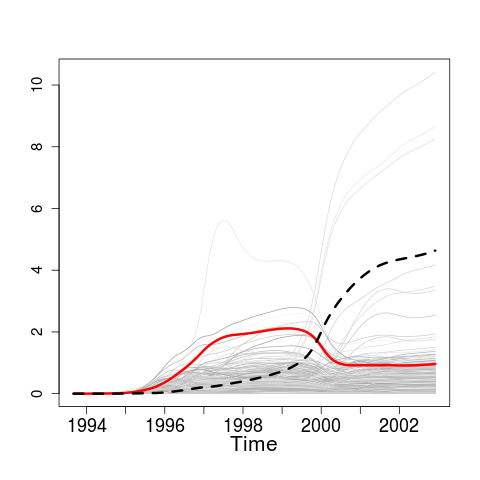}
&
\includegraphics[width=.5\columnwidth]{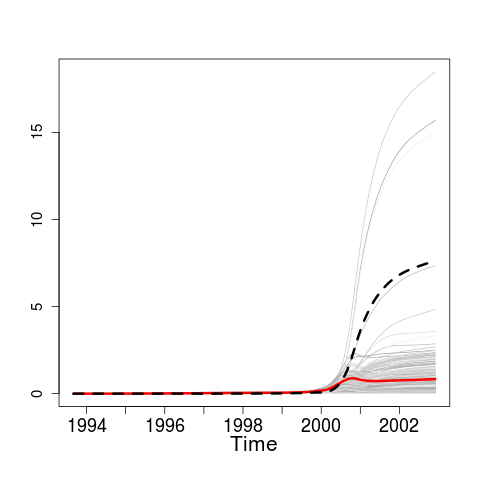}
&
\includegraphics[width=.5\columnwidth]{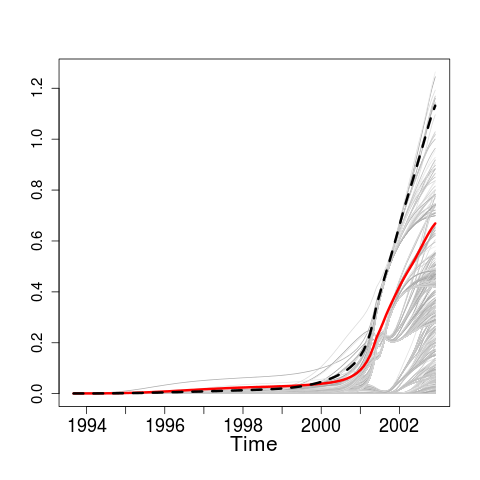}
\\
\end{tabular} 
\caption{(Color online) Fitted values for $U_t$ and $V_t$ for the
  arXiv data with $\lambda_t = 5$. Each light gray line corresponds to
  a paper (node) on the graph. The bold lines show the average of the
  10 papers with highest average $\hat{U}$ from 1996-1999, and 2000
  onwards (dashed). Each row in the heatmaps corresponds to a paper
  and time varies along the horizontal axis.}
\label{fig:arXiv}
\end{figure*}

To visualize how nodes in the network evolved, 
Fig.~\ref{fig:arXiv} displays results from the matrix factorization
model using a sequence of one-dimensional approximations ($K=1$). The
adjacency matrix is constructed so that $U_{t}$ scores nodes by their
importance to the average \emph{incoming} connections, and
$(U_t)_{1j}$ measures the time-varying authority of paper
$j$. $V_{t}$ yields similar scores based on outgoing connections. 
As
observed with the preferential attachment experiment, the paper
trajectories are smoothed effectively and important dynamics are
highlighted by employing penalties. Specifically, there are two important periods in
the data. The first period covers 1996-1999, and featured papers
mostly on an extension of string theory called M-theory. M-theory was
first proposed in 1995 and led to new research in theoretical
physics. A number of scientists, including Witten, Sen and Polchinski,
were important to the historical development of the theory, and as
seen in Tables~\ref{table:arXiv90s} and Table~\ref{table:arXiv00s},
our NMF approach identifies these important authors and their
works. From 1999-2000 the rate of citations to these papers tended to
decrease, while focus shifted to other topics and subfields that
M-theory gave rise to. These citation patterns are reflected in the bold and dashed 
trajectories in Fig.~\ref{fig:arXiv}. 
The displays of $V_t$ sparsity show 
that papers do not appear uniformly throughout time. Instead as other
network statistics show, papers `attach' at a faster rate around year
2000.

We provide comparisons with the alternative methodologies utilized in 
\cite{shedden-newman} to investigate dynamic citation network from the 
US Supreme Court. First, we apply the leading eigenvector modularity-based 
method for community discovery \cite{newman-modularity-matrix}
to the fully formed citation network ($t=112$). The second alternative 
methodology is a mixture model in \cite{shedden-newman} to extract groups 
of papers according to their common temporal citation profiles. 

\begin{figure}
\centerline{
\includegraphics[width=.5\columnwidth]{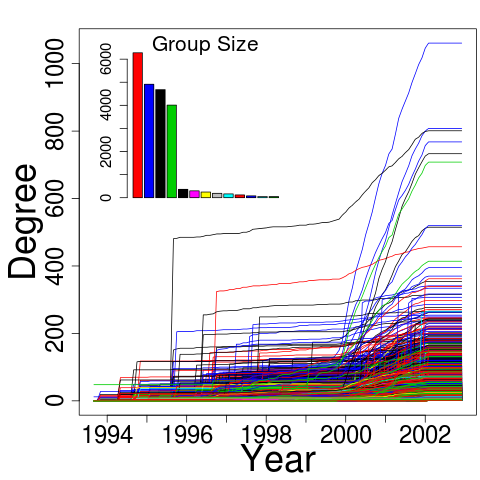} 
\includegraphics[width=.5\columnwidth]{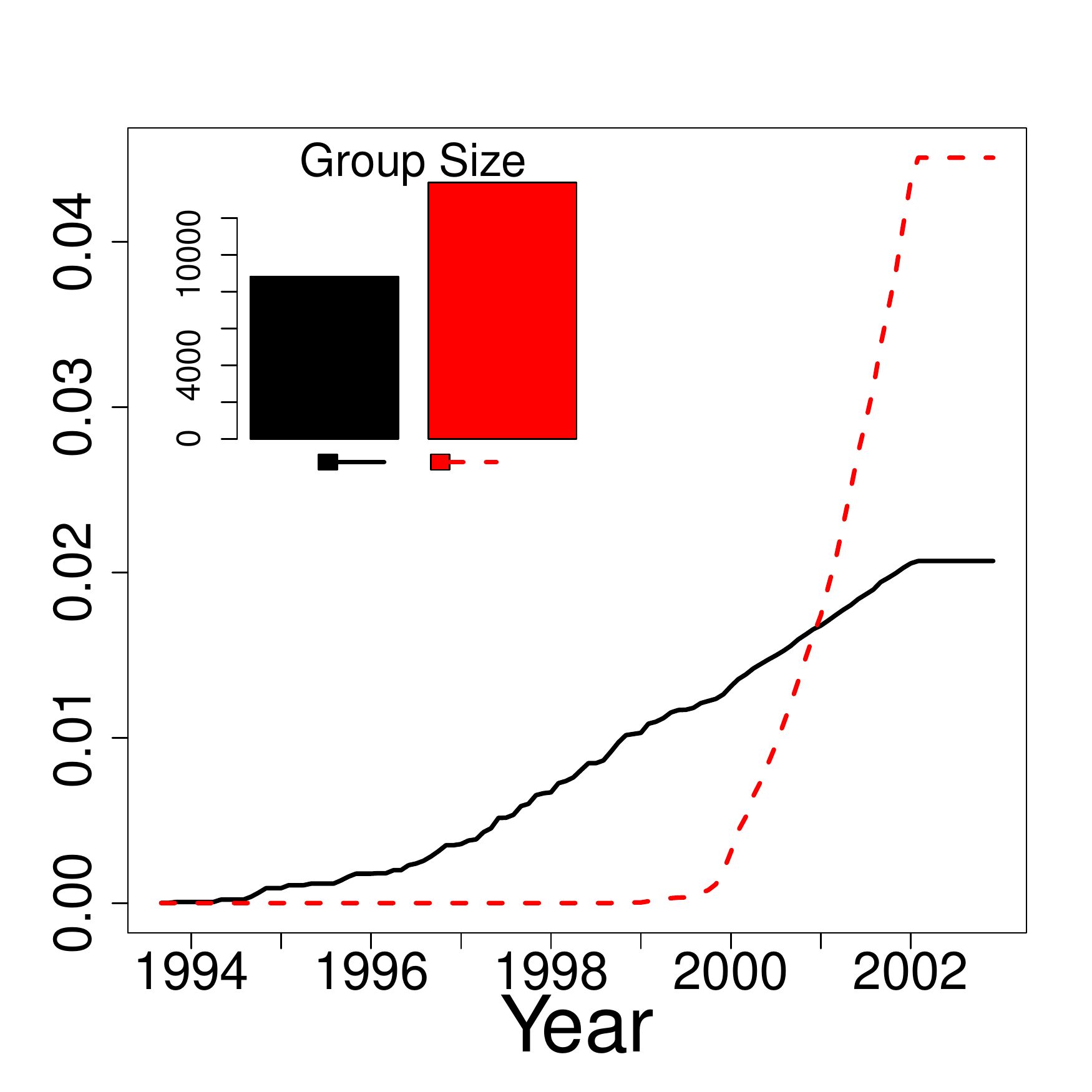} 
}
\caption{The left panel shows the degree of each node over all time points, colored by the leading eigenvector groupings. The right panel shows time-profiles for each group based on the mixture model of \cite{shedden-newman}.}
\label{fig:arXiv:alternatives}
\end{figure}

The left panel of Fig.~\ref{fig:arXiv:alternatives} shows 
the degree of each paper over time, colored by the leading 
eigenvector community assignments. The optimal number of
groups is over two hundred. There are four large groups of papers,
with the other groups containing only a handful of papers. 
This approach does not utilize the temporal profile of each paper, 
and as a consequence the groups are interpretable from a static 
connectivity point of view only.

The right panel of Fig.~\ref{fig:arXiv:alternatives} shows reasonable time-profiles 
from the mixture model. One group grows slowly 
from the beginning of the observational period, while the other group experiences rapid 
growth starting around the year 2000. 
These results compliment the NMF-based Fig.~\ref{fig:arXiv}, and together provide a robust 
methodology to identify important papers, as well as characterize the data in 
terms of the number and types of different nodes/groups in the data. 

\begin{table*}
\caption{The top 10 papers with highest average $\hat{U}$ from
  1996-1999. \# Citations counts all references to the work, including
  by papers outside of our data. These counts obtained via Google.}
\label{table:arXiv90s}
\centerline{
\begin{tabular}{llrrr}
\hline\hline
Title & Authors & In-Degree & Out-Degree & \# Citations (Google) \\ \hline
Heterotic and Type I String Dynamics from Eleven Dimensions & Horava and Witten & 783 & 18 & 2334 \\
Five-branes And $M$-Theory On An Orbifold & Witten & 169 & 15 & 251 \\
Type IIB Superstrings, BPS Monopoles, And Three- & Hanany and Witten & 437 & 20 & 844 \\
Dimensional Gauge Dynamics & & & & \\
D-Branes and Topological Field Theories &  Bershadsky, et al. & 271 & 15 & 463 \\
Lectures on Superstring and M Theory Dualities & Schwarz & 247 & 68 & 534 \\
D-Strings on D-Manifolds & Bershadsky et al. & 172 & 22 & 247\\
String Theory Dynamics In Various Dimensions & Witten & 263 & 0 & 2263 \\
Branes, Fluxes and Duality in M(atrix)-Theory & Ganor, et al. & 184 & 16 & 243\\
Dirichlet-Branes and Ramond-Ramond Charges & Polchinski & 370 & 0 & 2592\\
Matrix Description of M-theory on $T^5$ and $T^5/Z_2$ & Seiberg & 208 & 30 & 353 \\
\hline
\hline
\end{tabular}
}
\end{table*}

\begin{table*}
\caption{The top 10 papers with highest average $\hat{U}$ from 2000 onwards. }
\label{table:arXiv00s}
\centerline{
\begin{tabular}{llrrr}
\hline\hline
Title & Authors & In-Degree & Out-Degree  & \# Citations (Google)\\ \hline
The Large N Limit of Superconformal Field Theories and Supergravity & Maldacena & 1059 & 2 & 10697 \\
Anti De Sitter Space And Holography & Witten & 766 & 2 & 6956 \\
Gauge Theory Correlators from Non-Critical String Theory & Gubser et al. & 708 & 0 & 6004  \\
String Theory and Noncommutative Geometry & Seiberg and Witten & 796 & 12 & 3833 \\
Large N Field Theories, String Theory and Gravity & Aharony et a. & 446 & 74 & 3354 \\
An Alternative to Compactification & Randall and Sundrum & 733 & 0 & 5693\\
Noncommutative Geometry and Matrix Theory: Compactification on Tori & Connes et al. & 512 & 3 & 1810 \\
M Theory As A Matrix Model: A Conjecture & Banks et al. & 414 & 0 & 2460 \\
D-branes and the Noncommutative Torus & Douglas and Hull & 296 & 2 & 866 \\
Dirichlet-Branes and Ramond-Ramond Charges & Polchinski & 370 & 0 & 2592 \\
\hline
\hline
\end{tabular}
}
\end{table*}

\subsubsection{Global Trade Flows}
In this example, we analyze annual bilateral trade flows between 164
countries from 1980-1997 \cite{tradeFlows}. Thus, we observe a
dynamic, weighted graph at 18 time points, where each directional edge
denotes the total value of exports from one country to another. Since
trade flows can differ in size by orders of magnitude, we work with
trade values that are expressed in log dollars.

\begin{figure}
\centerline{
\includegraphics[width=.5\columnwidth]{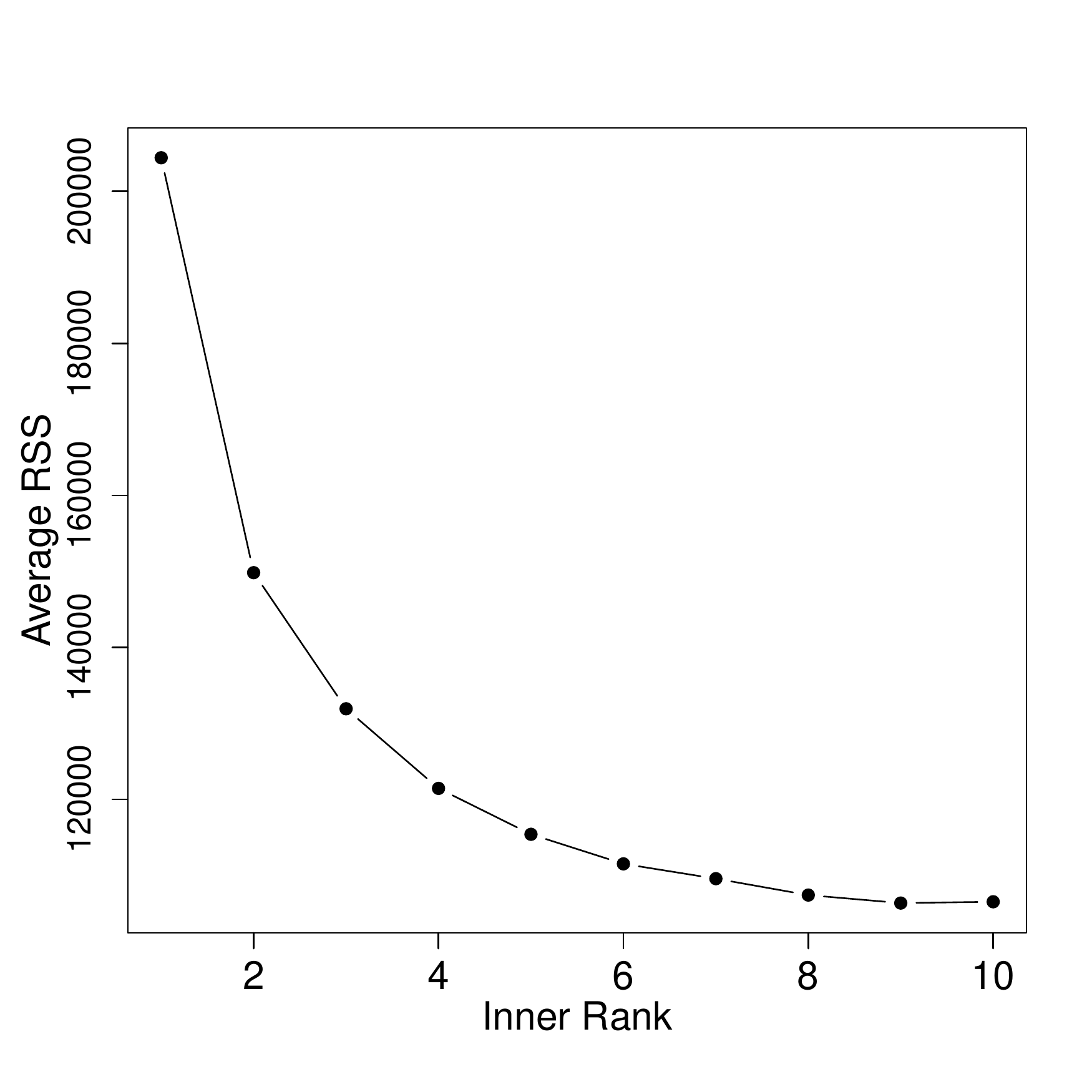}
\includegraphics[width=.5\columnwidth]{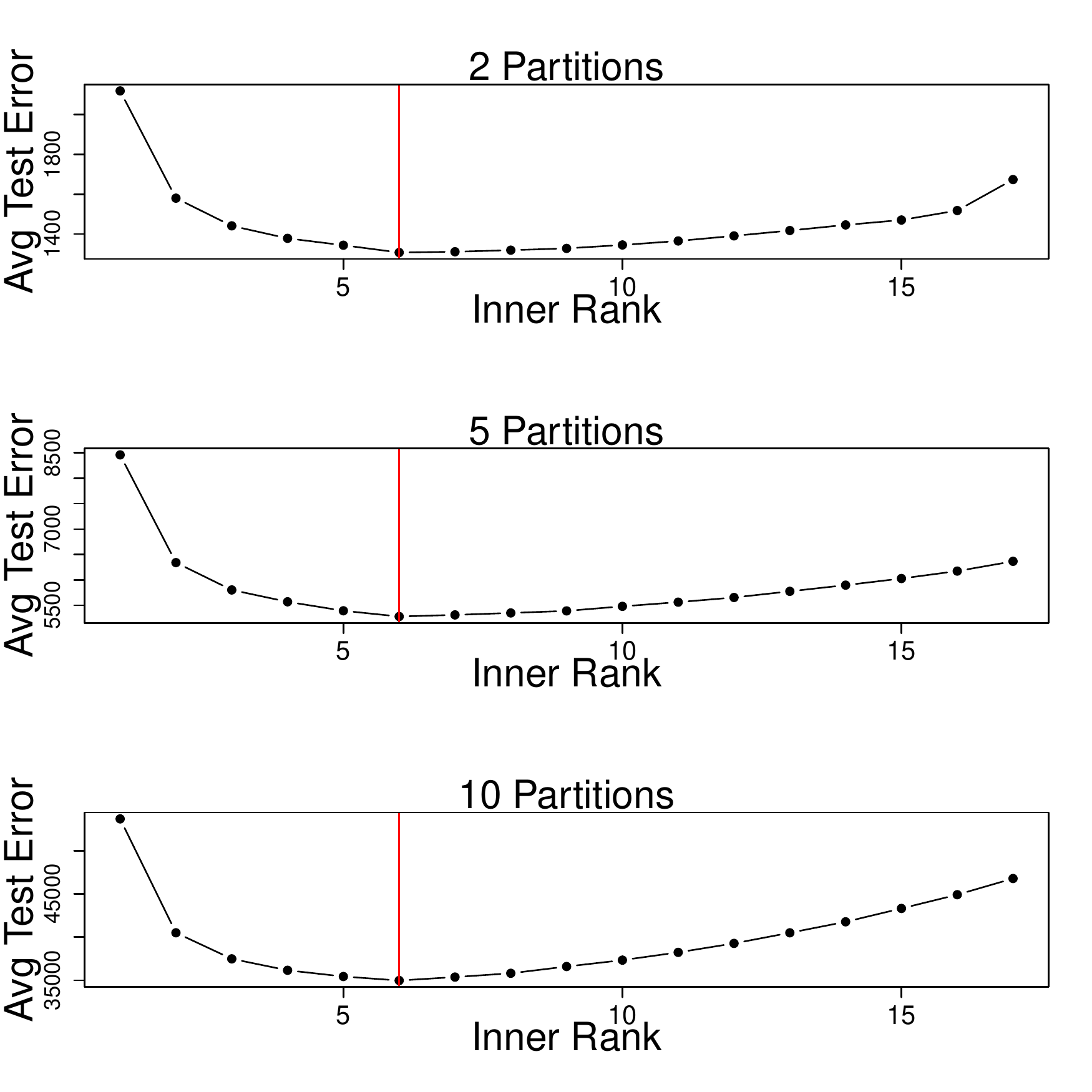}
}
\caption{Choosing $K$ for World Trade Data. The left panel shows the
  average residual sum of squares. The right panel shows the
  average test error obtained via cross validation for different number of partitions.  
  Cross validation consistently indicates 6
  communities ($K=6$) as optimal.}
\label{fig:worldtrade}
\end{figure}

\begin{figure*}
\begin{tabular}{ccc}
1980 & 1990 & 1997 \\
\includegraphics[width=.68\columnwidth, trim=3cm 3cm 1.25cm 2cm, clip=true]{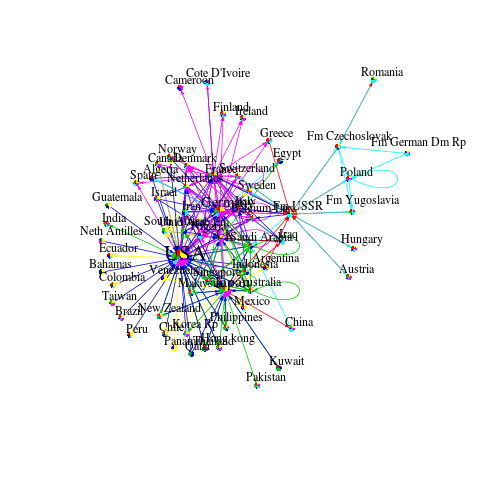} & 
\includegraphics[width=.68\columnwidth, trim=3cm 3cm 1.5cm 2cm, clip=true]{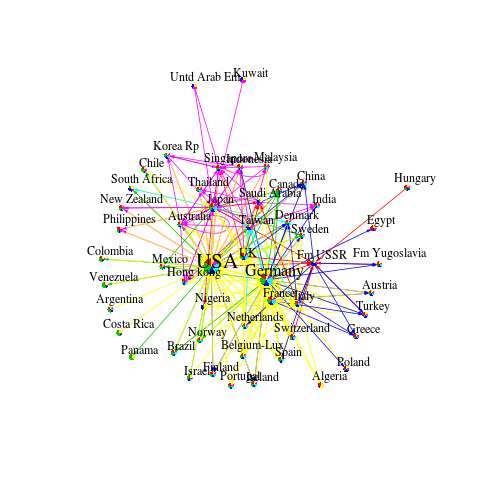} & 
\includegraphics[width=.68\columnwidth, trim=2.5cm 3cm 1.5cm 2cm, clip=true]{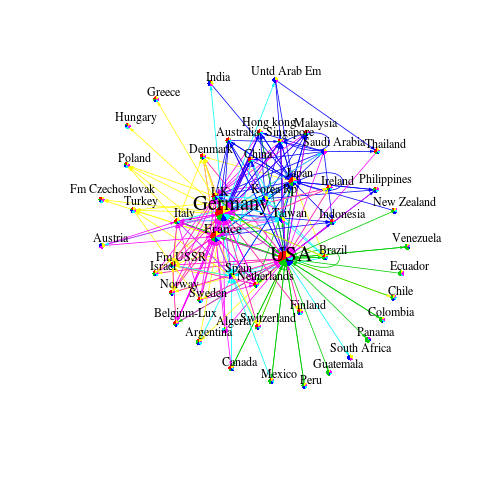} \\
\includegraphics[width=.68\columnwidth, trim=3cm 3cm 1.25cm 2cm, clip=true]{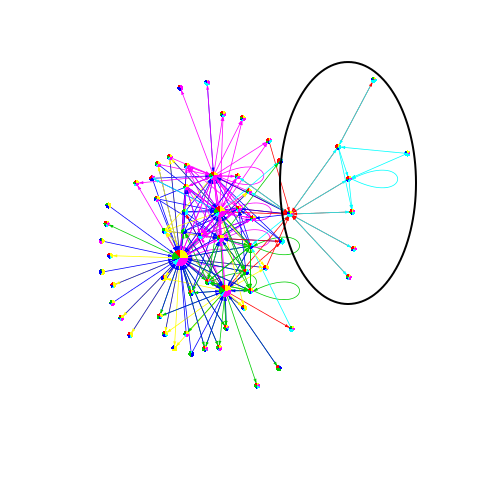} & 
\includegraphics[width=.68\columnwidth, trim=3cm 3cm 1.5cm 2cm, clip=true]{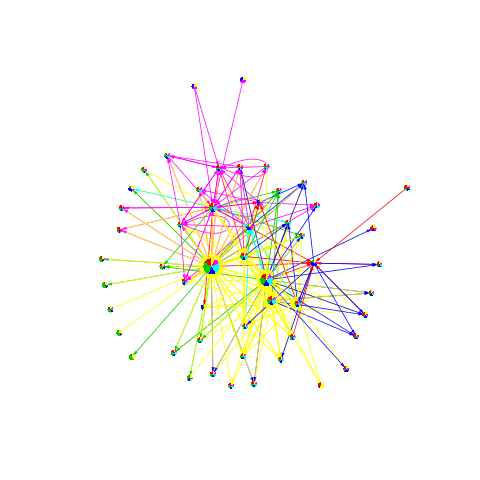} & 
\includegraphics[width=.68\columnwidth, trim=2.5cm 3cm 1.5cm 2cm, clip=true]{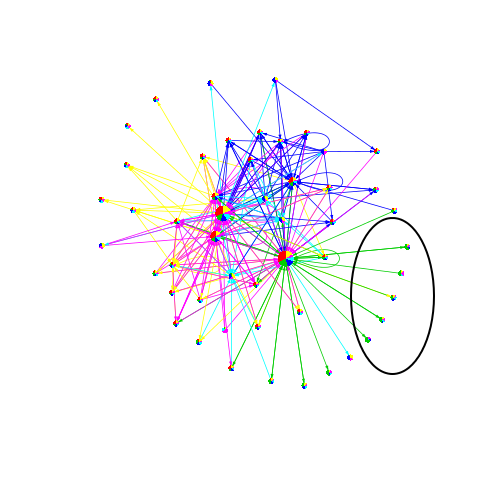} 
\end{tabular}
\caption{(Color online) World trade networks over time, where countries are
  colored corresponding to their membership in 6 communities. Edges are colored 
  by the community with largest relative contribution. 
  The bottom row shows the same network drawing without labels. }
\label{fig:worldMaps}
\end{figure*}

We fit a sequence of rank 6 NMFs, as identified in Fig.~\ref{fig:worldtrade}
through cross validation, and display the network based on fitted trade flows 
($\hat{A}_{t}=U_{t}V_{t}^{T}$) in Fig.~\ref{fig:worldMaps}. We show only three 
years (1980, 1990, 1997) due to space constraints. 

All
countries belong to more than one community, which reflects the interconnected nature 
of the global economy. However, there are countries, primarily from Africa and Central America, 
that are dominated by a single community or belong to only a subset of the six communities. For instance, 
in 1997, Ecuador, Venezuela and Panama only connect with the USA and hence, belong mostly to the green community. 

There are also interesting findings that correspond with historical events. 
For instance, in 1980 there is a strong 
community (circled in the figure) consisting of countries aligned with the former 
USSR, which acted as a hub. However by 1990, 
this community has dissolved, and is reflected in the edge and node colorings of these 
countries (more diversified 
trading relationships). In 1990, we also see the emergence 
of the so-called 'Asian miracles', countries in Asia that experienced persistent and
rapid economic growth in the 1990's \cite{asianMiracles2, asianMiracles}. These countries  
move closer to the center of the trading network with membership in all communities. 

\section{Discussion} \label{sec:discussion}
The main idea behind the approach presented in this paper is to abstract the network sequence to a sequence of lower dimensional spaces using matrix factorizations for visual exploration, community detection and structural discovery. Next, we highlight some of the strengths and weaknesses of this approach. 

\subsection{Strengths}
An important benefit is the versatility and scalability of matrix factorization model. Table \ref{table:runtimes} shows runtimes for all experiments. The computational cost is low enough to use in combination with other analysis and visualization tools. Moreover, the penalized NMF approach is compatible with both binary and weighted networks. 

\begin{table*}
\caption{Average runtimes for the penalized NMF with temporal and sparsity penalties. The computational time scales approximately linearly with the number of time points and nodes.}
\label{table:runtimes}
\centerline{
\begin{tabular}{ccccc}
\hline\hline
Data & Nodes & Time Points & Runtime (seconds) \\ \hline
Catalano & 400 & 7 & 0.29 \\
World Trade & 164 & 18 & 0.51 \\
Preferential Attachment & 10000 & 100 & 39.45\\
arXiv Citations & 22750 & 112 & 60.64\\
\hline
\hline
\end{tabular}
}
\end{table*}

Using the model as a basis for an exploratory visual tool can help users uncover different connectivity patterns and evolution in the data. The estimates of $U_{t}$ and $V_{t}$ can be used for community discovery or a ranking of nodes based on their importance to connectivity for subsequent analysis. Displays of the factorizations can provide a sense of the data complexity, namely the types and number of node evolutions.

\subsection{Weaknesses}
The optimal choice of tuning parameters ($\lambda_t, \lambda_s$) is dependent on perception and how the edge weights are scaled. This can limit the benefits of the proposed approach when given multiple datasets. 

Time plots and heatmaps to visualize each factor yield limited information about global topology. For example, one can see from Figs. \ref{fig:clustExpressionsPA} and \ref{fig:arXiv} that there are dominant nodes, but in principle, there could be many topologies that feature dominant nodes. One cannot say for sure without additional analysis that the networks follow a particular connectivity model. Thus, combining the matrix factorization model in this article with existing analysis and visualization tools can provide a more comprehensive analysis of the data.

\subsection{Future Work}
An important area of exploration would be to systematically compare penalized versions of NMF and SVD. In this work we chose to focus on NMF, since we find the corresponding displays preferable in terms of interpretability. This is generally consistent with existing literature on matrix factorization. However, SVD of graph related matrices have deep connections to classical spectral layout and problems in community detection. There may be classes of graph topologies and particular visualization goals under which SVD is preferable.

There could also be other types and combinations of penalties that are useful in visualization and detection of graph structure. For instance, depending on the precise meaning of a directional edge, one may desire both smoothness and sparsity for $U_{t}$, $V_{t}$ or both factors. Nonetheless, variants on the penalty structure will result in models that require roughly the same computational costs. Thus, this work provides evidence that penalized matrix factorization models are promising for structural and functional discovery in dynamic networks.

\bibliography{Visual}

\end{document}